\documentclass[seceq,preprint]{ptptex}

\usepackage{epsfig, color,graphicx}
\usepackage{amsmath}
\usepackage{amssymb}
\newcommand{\nn}{\nonumber \\}
\newcommand{\bea}{\begin{eqnarray}}
\newcommand{\ena}{\end{eqnarray}}
\newcommand{\vs}[1]{\vspace{#1 mm}}
\newcommand{\hs}[1]{\hspace{#1 mm}}
\renewcommand{\a}{\alpha}
\renewcommand{\b}{\beta}
\renewcommand{\c}{\gamma}
\renewcommand{\d}{\delta}
\newcommand{\e}{\epsilon}

\newcommand{\pa}{\partial}
\newcommand{\p}[1]{(\ref{#1})}
\newcommand{\tm}{\tilde{m}}
\newcommand{\tr}{\tilde{r}}
\newcommand{\squaret}{\kern1pt\vbox{\hrule height 0.9pt\hbox{\vrule width
0.9pt\hskip 2pt\vbox{\vskip 5.5pt}\hskip 3pt\vrule width 0.3pt}\hrule height
0.3pt}\kern1pt}
\begin{document}

\preprintnumber[3cm]{
KU-TP 023}
\title{\large Black Holes in the Dilatonic Einstein-Gauss-Bonnet Theory
in Various Dimensions I \\
-- Asymptotically Flat Black Holes --
}

\author{Zong-Kuan {\sc Guo},$^{a,}$\footnote{e-mail address: guozk at phys.kindai.ac.jp}
Nobuyoshi {\sc Ohta}$^{a,}$\footnote{e-mail address: ohtan at phys.kindai.ac.jp}
and Takashi {\sc Torii}$^{b,}$\footnote{e-mail address: torii at ge.oit.ac.jp}
}

\inst{
$^a$Department of Physics, Kinki University, Higashi-Osaka,
Osaka 577-8502, Japan\\
\vs{3}
$^b$ Department of General Education, Osaka Institute of Technology,
Asahi-ku, Osaka 535-8585, Japan
}
\abst{
We study spherically symmetric, asymptotically flat black hole solutions
in the low-energy effective heterotic string theory, which is the Einstein
gravity with Gauss-Bonnet term and the dilaton, in various dimensions.
We derive the field equations for suitable ansatz for general $D$ dimensions
and construct black hole solutions of various masses numerically in $D=4,5,6$
and 10 dimensional spacetime with $(D-2)$-dimensional hypersurface
with positive constant curvature.
A detailed comparison with the non-dilatonic solutions is made.
We also examine the thermodynamic properties of the solutions.
It is found that the dilaton has significant effects
on the black hole solutions, and we discuss physical consequences.
}

\maketitle

\section{Introduction}

One of the most important problems in theoretical physics is the formulation of
the quantum theory of gravity and its application to physical system to
understand physics at strong gravity. The leading candidates for that including
all the fundamental forces of elementary particles are the ten-dimensional
superstring theories. The area where such quantum gravity plays the significant
role includes the cosmology and black hole physics.
There has been interest in the application of string theory to these subjects.
Since it is still difficult to study geometrical settings in superstring theories,
most analyses have been performed by using low-energy effective theories
inspired by string theories. These effective theories are the supergravities
which typically involve not only the metric but also the dilaton field
(as well as several gauge fields).

The first attempt at understanding black holes in the Einstein-Maxwell-dilaton
system was made in Refs.~\citen{GM,GHS}, in which a static spherically symmetric
black hole solution with dilaton hair was found. After this, many solutions
were discussed in various models. On the other hand,
it is known that there are correction terms of higher orders in the curvature to the
lowest effective supergravity action coming from superstrings.
The simplest correction is the  Gauss-Bonnet (GB) term coupled to the dilaton field
in the low-energy effective heterotic string~\cite{gro87}.
(We ignore other gauge fields and forms for simplicity.)
It is then natural to ask how the black hole solutions are affected by
the higher order terms in these effective theories.

When the dilaton is dropped or is set to a constant, this correction is known
as the first term except for the cosmological constant and Einstein-Hilbert
action in the Lovelock gravity~\cite{Lovelock,Zumino} which is the most
general metric theory of gravity yielding conserved second order equations of
motion in arbitrary number of dimensions $D$.
It is a natural generalization of Einstein's general relativity (GR) to higher
dimensions without ghost, and for this reason it was conjectured~\cite{zwi85}
and indeed found to be the low-energy effective theory of strings.
Motivated by these observations, there have been many works on the black hole
solutions in the Lovelock theories~\cite{GG,CNO,TM,mae07}.
In the four-dimensional spacetime,
the GB term does not give any contribution because it becomes a surface term
and gives a topological invariant. Boulware and Deser~\cite{sil03}
discovered a static, spherically symmetric black hole solutions of such
models in more than four dimensions. In the system with a negative cosmological
constant, black holes can have horizons with nonspherical topology such as
torus, hyperboloid, and other compactified submanifolds. These
solutions were originally found in general relativity and are
called topological black holes~\cite{bro84}. Topological black
hole solutions were studied in GB gravity~\cite{Cai}.
It is of interest to see how these are modified by the presence of a dilaton.

Another motivation for this work is the following. There has been recently
a renewed interest in these solutions as the application to the
calculation of shear viscosity in strongly coupled gauge theories using black hole
solutions in five-dimensional Einstein-GB theory via AdS/CFT correspondence~\cite{vis}.
Almost all these studies consider a pure GB term without a dilaton,
or assume a constant dilaton, which is not a solution of the heterotic string.
It is, however, expected that AdS/CFT correspondence is valid within
the effective theories of superstring. It is thus again important
to investigate how the properties of black holes
are modified when the dilaton is present. The inclusion of the
dilaton was also considered by Boulware and Deser~\cite{bou86},
but exact black hole solutions and their thermodynamic properties
were not discussed. Callan {\it et al.}~\cite{cal88} considered
black hole solutions in the theory with a higher-curvature term
$R_{\mu\nu\rho\sigma}R^{\mu\nu\rho\sigma}$ and the dilaton field,
and Refs.~\citen{mig93,TYM} took both the GB term and the dilaton
field into account in four-dimensional spacetime.

This is the first of a series of papers in which we study black holes
in the dilatonic Einstein-GB theory in $D$ dimensions
with $(D-2)$-dimensional Euclidean  manifold with constant curvature
of signature $k=\pm 1,0$.
In this paper, we consider asymptotically flat black hole solutions
in static spherically symmetric ($k=1$) spacetime and discuss their thermodynamic
properties.
In the subsequent publications, we intend to study $k=0, -1$ topological black
hole solutions with possible cosmological constant including asymptotically
anti-de Sitter solutions, which should be useful for studying the dynamics of
strongly coupled gauge theories.

This paper is organized as follows. In \S~2, we first
introduce the action of the dilatonic Einstein-GB
theory, and our metric for $D$-dimensional spacetime. We then
summarize the field equations. In \S~3, we focus on an asymptotically
flat Schwarzschild-type metric and impose the boundary conditions for
the metric functions and the dilaton field on the event horizon and at
spatial infinity.

In \S~4 through \S~8, we present the black hole solutions
in the cases of $D=4, 5, 6$ and $10$.
For comparison, we first summarize the non-dilatonic solution in
Einstein-GB theory in various dimensions in \S~4. Then, in \S~5, we discuss $D=4$
black hole solutions for the dilaton coupling $\c=1$ and recover the results
in Ref.~\citen{TYM} in subsection~5.1. In subsection~5.2, we discuss the solutions
for the dilaton coupling $\c=\frac12$ which is the value adopted mainly in
this paper, and find that there is one notable difference from the other dilaton
couplings. That is, there are two solutions for certain range of mass parameters
at the smaller values for the dilaton coupling $\c=1$, but there is not such
region for the coupling $\c=\frac12$. We find that there is no other big
difference between these cases in the behaviors of the metrics and dilaton fields.
In particular, in both cases, there is a lower bound on the mass and horizon
for the solutions to exist.
On the other hand, we find that $D=5$ solutions exhibit quite different
properties, e.g. they exist even in the vanishing limit of the horizon radius
in contrast to four dimensions. This is presented in \S~6.
We find that properties of the solution in $D=6$ through $D=10$ are very similar so that
we present results only for $D=6$ and 10 cases in \S~7 and \S~8, respectively.
In these sections, we also make detailed comparison with the non-dilatonic case.

In \S~9, we investigate the thermodynamic properties of the dilatonic black
holes with the GB term.
Here again, we find that the five-dimensional black holes
have quite distinctive properties from other-dimensional solutions.
The heat capacity of the non-dilatonic black hole is negative for large mass
but changes sign to be positive for smaller mass in five-dimensional solutions.
However, it is always negative as the mass is varied if  the dilaton field is added.
In dimensions other than five, the heat capacity is always negative for both
the dilatonic and the non-dilatonic black holes.
Section~10 is devoted to conclusions.

\section{Dilatonic Einstein-GB theory}

We consider the following low-energy effective action for the heterotic string:
\bea 
S=\frac{1}{2\kappa_D^2}\int d^Dx \sqrt{-g} \left[R - \frac12
 (\partial_\mu \phi)^2
 + \a_2 e^{-\c \phi} R^2_{\rm GB} \right],
\label{act}
\ena 
where $R$ is the scalar curvature, $\phi$ a dilaton field,
$R^2_{\rm GB} = R_{\mu\nu\rho\sigma} R^{\mu\nu\rho\sigma} - 4 R_{\mu\nu} R^{\mu\nu}
+ R^2$ the GB combination, $\kappa_D^2 =8\pi G_D$
a $D$-dimensional gravitational constant, and $\alpha_2=\a'/8\;(> 0)$ is a numerical
coefficient given in terms of the Regge slope parameter $\alpha'$.

Here $\c$ is the coupling constant of the dilaton field.
There is ambiguity in the choice of this constant.
If we first make dimensional reduction of the system in the string frame
to $D$ dimensions and then change to the Einstein frame, we would get
$\c=\sqrt{2/(D-2)}$.
This is the choice, for example, $\c=1$ in Ref.~\citen{TYM} for $D=4$.
However, if we first go to the Einstein frame in ten dimensions with $\c=\frac12$
and then make the dimensional reduction, we get the value $\c=\frac12$ in any dimensions.
Both choices have their own right, but it is convenient to take the same value
for our analysis of black holes in any dimensions.
So in this paper we take the second viewpoint and choose $\gamma=\frac12$ mainly
in our following study of the black hole solutions.
In order to see how the results depend on this choice, however,
we also examine the black hole solutions for both the values of $\c=1$
and $\frac12$ in four dimensions in \S~5. This also serves to check
the consistency of our results with Ref.~\citen{TYM}.
In fact we will find that the solutions exhibit quite similar behaviors
although there is some difference in details. In order to make our formulae
valid for any case, we will keep $\c$ wherever possible.

Varying the action~\p{act} with respect to $g_{\mu\nu}$, we obtain
the gravitational equation:
\begin{equation} 
\label{GB-eq}
G_{\mu\nu}
-\frac12\biggl[\nabla_{\mu}\phi\nabla_{\nu} \phi -\frac12 g_{\mu\nu}(\nabla\phi)^2\biggr]
+\alpha_2 e^{-\gamma\phi}\Bigl[H_{\mu\nu}
+4(\gamma^2\nabla^{\rho}\phi\nabla^{\sigma}\phi
-\gamma\nabla^{\rho}\nabla^{\sigma}\phi)P_{\mu\rho\nu\sigma}\Bigr]
=0,
\end{equation} 
where
\begin{eqnarray} 
&&G_{\mu\nu}\equiv R_{\mu\nu}-{1\over 2}g_{\mu\nu}R,
\\
&&H_{\mu\nu}\equiv 2\Bigl[RR_{\mu\nu}-2R_{\mu \rho}R^{\rho}_{~\nu}
-2R^{\rho\sigma}R_{\mu\rho\nu\sigma}
+R_{\mu}^{~\rho\sigma\lambda}R_{\nu\rho\sigma\lambda}\Bigr]
-{1\over 2}g_{\mu\nu}R^2_{\rm GB},
\\
&& P_{\mu\nu\rho\sigma}\equiv
R_{\mu\nu\rho\sigma}+2g_{\mu[\sigma}R_{\rho]\nu}
+2g_{\nu[\rho}R_{\sigma]\mu} +Rg_{\mu[\rho}g_{\sigma]\nu}.
\label{EGB:eq}
\end{eqnarray} 
$P_{\mu\nu\rho\sigma}$ is the divergence free part of the Riemann tensor,
i.e.
\begin{equation} 
\nabla_\mu P^{\mu}_{~\nu\rho\sigma}=0.
\end{equation} 
The equation of the dilaton field is
\begin{eqnarray} 
\label{dil-eq}
\squaret \phi -\alpha_2 \gamma e^{-\gamma\phi}  R^2_{\rm GB}=0,
\end{eqnarray} 
where $\squaret$ is the $D$-dimensional d'Alembertian.

To derive black hole solutions in this system, let us consider
the  line element in $D$-dimensional static spacetime
\bea 
ds^2=-e^{2u(r)}dt^2 + e^{2v(r)}dr^2 + r^2 h_{ij}dx^i dx^j \,,
\ena 
where $h_{ij}dx^i dx^j$ represents the line element of a
$(D-2)$-dimensional hypersurface with constant curvature of signature
$k$ and volume $\Sigma_k$ for $k=\pm 1,0$.
We consider $k=1$ for the black hole solutions in this paper,
but keep $k$ wherever possible.
The solutions with $k=1$ is spherically symmetric in four-dimensional spacetime,
but the $(D-2)$-dimensional hypersurface can have rich structure and is not
necessarily homogeneous~\cite{Wolf,3-einstein_2,3-einstein_1}.

Our basic equations then give~\footnote
{The system~\p{act} was considered in Ref.~\citen{BGO} for application to
cosmological model with accelerating expansion. Time-dependent solutions
for $p$- and $q$-dimensional external and internal spaces were studied.
The field equations can be derived from the results given there by the replacement
\begin{eqnarray*}
&&
t \to -ir, \quad
ds_p^2 \to -dt^2, \quad
p=1, \quad
\sigma_p=0, \quad
q=D-2, \quad
\sigma_q=k,
 \nn &&
u_0 \to v(r), \quad
u_1 \to u(r), \quad
u_2 \to \ln r.
\end{eqnarray*}
}
\bea 
\label{1}
&&F \equiv (D-2)_3 A(r) -2(D-2) \frac{\dot u}{r} +\frac12 \dot\phi^2
+ \a_2 e^{-2v-\c\phi} (D-2)_3 \Big[ (D-4)_5 A^2(r) \nn
&& \hs{10} - 4(D-4) A(r) \frac{\dot u-\c \dot\phi}{r}
+ 4\c \frac{e^{2v}}{r^2}(k-3e^{-2v})\dot u\dot \phi \Big] =0, \\
\label{2}
&&G \equiv (D-2)_3 A(r) + 2(D-2) \frac{\dot v}{r} - \frac12 \dot\phi^2
+ \a_2 e^{-2v-\c\phi} (D-2)_3 \Bigg[ (D-4)_5 A^2(r) \nn
&& \hs{10}+ 4(D-4) A(r) \frac{\dot v}{r}
+ 4\c \Big\{ \ddot\phi-\Big(\dot v+\c\dot \phi-\frac{D-4}{r}\Big)\dot\phi \Big\} A(r)
+8 \c\frac{\dot v\dot\phi}{r^2}\Bigg] =0, \\
&&H \equiv (D-3)_4 A(r) - 2(D-3) \frac{\dot u-\dot v}{r} - \frac12 \dot\phi^2 -2U(r) \nn
&&
\hs{10}+ \a_2 e^{-2v-\c\phi} (D-3) \Bigg[ (D-4)_6 A^2(r) - 4(D-4)_5 A(r)
\frac{\dot u-\dot v-\c\dot\phi}{r}
\nn
&& \hs{10} - 4(D-4)A(r) \Big\{ \c^2\dot\phi^2 + U(r)
 -\c (\ddot\phi+(\dot u-\dot v)\dot\phi) \Big\} \nn
&& \hs{10}+8 \frac{\c \dot\phi}{r}\dot u \Big(2\dot v+\c\dot\phi-\frac{D-4}{r} \Big)
-8 (D-4) \frac{\dot v}{r^2}(\dot u-\c\dot\phi)
-8\c \frac{\dot u \ddot\phi+ \dot\phi U(r)}{r}\Bigg] =0,
\label{3}
\\
\label{4}
&&F_\phi \equiv
 \ddot\phi+\dot\phi(\dot u-\dot v)+(D-2)\frac{\dot\phi}{r}
-\a_2\c e^{2v-\c\phi}R_{\rm GB}^2=0,
\ena 
where the dot denotes the derivative with respect to $r$,
and we have defined
\bea 
&&(D-m)_n \equiv (D-m)(D-m-1)(D-m-2)\cdots(D-n), \\
&&U(r) \equiv \ddot u + {\dot u}^2 - \dot u \dot v, \\
&&A(r) \equiv \frac{e^{2v}}{r^2} (k-e^{-2v}),
\ena 
and the GB term is expressed as
\bea 
R^2_{\rm GB} = (D-2)_3 e^{-4v} \biggl[
(D-4)_5 A^2(r) - 8 \frac{\dot u \dot v}{r^2}
- 4 A(r)\left\{ U(r)+\frac{(D-4)(\dot u-\dot v)}{r}\right\} \biggr].~~
\ena 
Eqs.~\p{1}--\p{4} are not all independent but satisfy
\bea 
\dot F + \Big(\dot u-2\dot v+\frac{D-2}{r} \Big) F -\dot u G
-\frac{D-2}{r}H - \dot\phi F_\phi=0.
\ena 
This serves to check the consistency of the results.

\section{Metrics and boundary conditions}

\subsection{Basic equations}

In order to discuss asymptotically flat solutions, we parametrize the metric as
\bea 
\label{metric}
ds_D^2 = - \left(k-\frac{2Gm}{r^{D-3}} \right) e^{-2\d} dt^2
+ \left(k-\frac{2Gm}{r^{D-3}} \right)^{-1} dr^2 + r^2 h_{ij}dx^i dx^j.
\ena 
The mass function $m=m(r)$ and the lapse function $\d=\d(r)$ depend only
on the radial coordinate $r$. The field equations~\p{1}, \p{2} and \p{4} then give
\bea 
&& \tm' \frac{D-2}{\tr^{D-4}}h -\frac14 B \tr^2 {\phi'}^2
 - \frac{1}{2}(D-1)_4\,e^{-\c\phi}\frac{(k-B)^2}{\tr^2} \nn
&& \hs{10} + 2(D-2)_3\, \c e^{-\c\phi}B(k-B)(\phi''-\c {\phi'}^2) \nn
&& \hs{10} + (D-2)_3\,\c e^{-\c\phi}\phi'\frac{(k-B)[(D-3)k-(D-1)B]}{\tr}=0\,,
\label{g1} \\
&& \delta'(D-2)\tr h + \frac12 \tr^2 {\phi'}^2
 -2(D-2)_3\, \c e^{-\c\phi}(k-B)(\phi''-\c {\phi'}^2) =0 \,,
\label{g2} \\
&&
(e^{-\d} \tr^{D-2} B \phi')' = \c (D-2)_3 e^{-\c\phi-\d} \tr^{D-4}
\Big[ (D-4)_5 \frac{(k-B)^2}{\tr^2} + 2(B'-2\d' B)B' \nn
&& \hs{10} -4(k-B)BU(r)
-4\frac{D-4}{\tr}(B'-\d'B)(k-B) \Big],
\label{g3}
\ena 
where we have defined the dimensionless variables
\bea
\tr \equiv \frac{r}{\sqrt{\a_2}}, ~~
\tm \equiv \frac{Gm}{\a_2^{(D-3)/2}},
\ena
and the prime in the field equations
denotes the derivative with respect to $\tr$. Here we have also defined
\bea 
B &\equiv&  k-\frac{2\tm}{\tr^{D-3}}, \\
h &\equiv& 1+2(D-3) e^{-\c\phi} \Big[ (D-4) \frac{k-B}{\tr^2}
 + \c \phi'\frac{3B-k}{\tr}\Big], \\
\tilde h &\equiv& 1+2(D-3) e^{-\c\phi} \Big[(D-4)\frac{k-B}{\tr^2}
+\c\phi'\frac{2B}{\tr} \Big],
\label{ht} \\
U(r) &\equiv& \frac{1}{2 \tilde h} \Biggl[ (D-3)_4 \frac{k-B}{\tr^2 B}
-2\frac{D-3}{\tr}\Big(\frac{B'}{B}-\d'\Big) -\frac12 \phi'^2 \nn
&& + (D-3)e^{-\c\phi} \biggl[ (D-4)_6 \frac{(k-B)^2}{\tr^4 B}
- 4 (D-4)_5 \frac{k-B}{\tr^3}\Big(\frac{B'}{B}-\d'-\c\phi'\Big) \nn
&& -4(D-4)\c \frac{k-B}{\tr^2}\Big( \c \phi'^2 +\frac{D-2}{\tr}\phi'-\Phi \Big)
+8 \frac{\c\phi'}{\tr} \Big\{\Big(\frac{B'}{2}-\d' B\Big)\Big(\c\phi'-\d'
+\frac{2}{\tr} \Big) \nn
&& \hs{20} -\frac{D-4}{2\tr}B' \Big\}
+4(D-4)\Big(\frac{B'}{2B}-\d' \Big)\frac{B'}{\tr^2}-4\frac{\c}{\tr}\Phi (B'-2\d'B)\biggr]
\Bigg], \\
\Phi &\equiv& \phi'' +\Big(\frac{B'}{B}-\d' +\frac{D-2}{\tr}\Big) \phi'.
\ena 
The function $U(r)$ here is obtained by solving the field equation~\p{3}.

Note that these equations have a symmetry under
\bea 
\phi \to \phi-\phi_\infty, ~~
\tr \to e^{\frac{1}{2}\c\phi_\infty} \tr, ~~
\d \to \d, ~~
\tilde m \to e^{\frac{D-3}{2}\c\phi_\infty} \tilde m.
\label{shift}
\ena 
This can be used to shift the asymptotic value of the dilaton field to zero
even if we compute metric functions and the dilaton field for given boundary conditions
at the horizon. Since the spacetime is time independent, there is another
shift symmetry under
\bea 
\delta \to \delta - \delta_{\infty}, ~~
t \to  e^{-\delta_\infty}t,
\label{shift2}
\ena 
which may be used to shift the asymptotic value of $\delta$ to zero.

\subsection{Boundary conditions}

{}From now on, we consider $k=1$.
The horizon is defined by the condition $B(\tilde r_H)=0$, i.e.
\bea 
\tilde m(\tr_H) = \frac{\tr_H^{D-3}}{2}.
\ena 
In what follows, the quantities evaluated at the horizon are
denoted with subscript $H$.
We impose the following boundary conditions for the metric functions and
the dilaton field at the event horizon and spatial infinity:
\begin{enumerate}
\item
Asymptotic flatness at spatial infinity ($\tr \to \infty$):
\bea 
\label{mass}
\tm(\tr) &\to& \tilde M< \infty, \\
\d(\tr) &\to& 0, \\
\phi(\tr) &\to& 0.
\ena 
\item
The existence of a regular horizon $\tr_H$:\footnote{
There is a case where the metric functions and curvatures of
spacetime are finite although the scalar field diverges at
the horizon~\cite{singluarBH,singluarBH2}. It is shown, however,
that such solution has unphysical properties~\cite{Xanthopoulos,Sudarsky}.
Hence we assume the dilaton field is finite at and outside the horizon.}
\bea 
\label{hor}
2 \tm_H &=& \tilde r_H^{D-3}, \\
|\d_H| &<& \infty, \\
|\phi_H| &<& \infty.
\ena 
\item
The event horizon is the outermost one and the regularity of spacetime for $\tr > \tr_H$:
\bea 
2\tm(\tr) &<& \tr^{D-3}, \\
|\d(\tr)| &<& \infty, \\
|\phi(\tr)| &<& \infty.
\ena 
\end{enumerate}

Looking at Eq.~\p{g3} for the dilaton, we see that it appears singular at the horizon
$B=0$ if we solve for $\phi''$. In order to deal with this, we expand the equations
and field variables in the power series of $\tilde{r}-\tilde{r}_H$ to guarantee
the regularity at the horizon. From the zeroth order term in $\tilde{r}-\tilde{r}_H$
of Eq.~\p{g1}, we find
\bea 
hB'_H = \frac{D-3}{\tr_H}+e^{-\c\phi_H}\frac{(D-3)_5}{\tr_H^3}
= \frac{(D-5)\tilde h_H +D-1}{2\tr_H},
\ena 
where we have used Eq.~\p{ht} in deriving the second equality.
Using this in Eq.~\p{g3} at the horizon, we obtain the quadratic equation
determining $\phi_H'$:
\bea 
&& \hs{-7} 2  C \gamma  \Big[2 (D-3) +(D-4) (3 D-11) C  \nn
&&~~~~ +(D-4) C^2 \Big\{(D-4)_5 +(D-2) (3 D-11) \gamma ^2\Big\}
+2 (D-2)_5\, C^3 \gamma ^2\Big] \tr_H^2 \phi_H'^2 \nn
&&+2\Big[(D-1)_2\,(D-4) C^2 \Big\{2 +2 C  -(D-4)_5\, C^2\Big\} \gamma ^2  \nn
&&~~~~ -  \{1+(D-4) C\}^2\, \{2 (D-3)+ (D-4)_5\, C\} \Big]\tr_H \phi_H' \nn
&& +(D-1)_2\, C \Big[2 (D-2) -4 (D-4) C -(D-4)^2 (D+1) C^2\Big]\gamma=0,
\label{phiprime}
\ena 
where we have defined
\bea
C=\frac{2(D-3)e^{-\c\phi_H}}{\tr_H^2}.
\label{c}
\ena
Eq.~\p{phiprime} shows that the values of the dilaton field and its derivatives
are related, and $\phi'_H$ is obtained in terms of $\phi_H$.
{}From the first-order terms of $\tilde{r}-\tilde{r}_H$, we can express
the second derivative $\phi''_H$ by $\phi_H$ and $\phi'_H$, and use
their analytic solution for the first step of integration.

In the asymptotic region far from the horizon, the curvature of the spacetime
is small and the GB term is negligible. The dilaton field then behaves as
\bea 
\phi \sim -\frac{\Sigma}{\tr^{D-3}},
\ena 
where $\Sigma$ is the dilaton charge. This global charge is not a free parameter
of the solution but is fixed by the mass of the black hole.
In this sense, the dilaton charge is classified into the secondary hair.

\section{Non-dilatonic black hole solutions}

It will be instructive to compare our results with the non-dilatonic case.
Let us derive physical quantities for this case first. When the dilaton
field is absent (i.e., Einstein-GB system), we substitute $\phi\equiv 0$
and $\gamma=0$ into Eqs.~\p{g1} and \p{g2}. In the $D=4$ case, the GB
term is total divergence and does not give any contribution to the field
equations. As a result, the solution reduces to the Schwarzschild solution.

For the $D\geq 5$ case, the field equations can be integrated to yield~\cite{sil03,TM}
\bea 
\bar{B}=1-\frac{2\bar{m}}{\tr^{D-3}},
\ena 
\bea 
\delta = 0,
\ena 
where
\bea 
&&\bar{m}=\frac{\tilde{r}^{D-1}}{4(D-3)_4}
\Biggl[-1 \pm \sqrt{1+\frac{8(D-3)_4\bar{M}}{\tilde{r}^{D-1}}} \Biggr],
\label{ND-M}
\ena 
and $\bar{M}$ is an integration constant corresponding to the asymptotic value
$\bar{m} (\infty)$ for the plus sign in Eq.~\p{ND-M}.
Throughout this paper, all the quantities with a bar denote those of
the non-dilatonic solution normalized by $\alpha_2$ as in the dilatonic case.
In the $\alpha_2\to 0$ limit, the solutions with the plus sign approach
the Schwarzschild solutions. This means that they can be considered to be
the solution with GB correction to GR. On the other hand,
the solutions with the minus sign do not have such a limit.
For these reasons, we call the solutions with plus (minus) sign the (non-)GR branch.

For $\bar{M}=0$, the metric function becomes
\bea 
\label{metric-f1}
\bar{B}=\left\{
\begin{array}{ll}
1& (\mbox{GR branch}) \\
1+\dfrac{\tr^2}{\tilde{\ell}_{\rm eff}^2}  & (\mbox{non-GR branch})
\end{array}
\right.
\ena 
where $\tilde{\ell}_{\rm eff}^2=(D-3)_4$. Hence the spacetime is Minkowski
in the GR branch while the spacetime is anti-de Sitter in the non-GR
branch although the pure cosmological constant $\Lambda$ is absent.

For $\bar{M}\ne 0$, besides the central singularity at $\tr=0$, there can be
another known as the branch singularity at finite radius $\tr_b>0$,
which is obtained by the condition that the inside of the square root
in Eq.~(\ref{ND-M}) vanishes. We find the $\bar{M}$-$\tr_b$ relation
\begin{equation} 
\label{branch-sing}
\bar{M}=-\frac{\tr_b^{D-1}}{8(D-3)_4}.
\end{equation}   
This implies that the branch singularity appears for negative mass parameter.

It can be shown that there is no black hole solution in the non-GR branch.
On the other hand, in the GR branch, Eq.~\p{ND-M} evaluated at the horizon
$\bar B=0$ gives
\begin{equation} 
\bar{M}=\frac12 \tilde{r}_H^{D-5} \Bigl[ \tilde{r}_H^2 +(D-3)_4\Bigr].
\label{nondil}
\end{equation} 
This is the $\bar{M}$-$\tr_H$ relation for the black hole
without the dilaton field.

This relation indicates that the $D=5$ case is qualitatively different from
other dimensional solutions. In four dimensions, the constant term in
Eq.~\p{nondil} vanishes and the relation between $\bar M$ and $\tr_H$ is linear,
and it vanishes in the $\tr_H \to 0$ limit. In dimensions higher than five,
the constant remains but the overall factor of $\tr_H$ has positive power,
and it also vanishes as the size of black hole becomes zero.
In the five-dimensional case, however, $\bar{M}$ takes
nonzero finite value ($\bar{M}=1$) in the zero horizon-radius limit. Since the
derivative of the metric function $\bar{B}$ of such tiny black hole goes to zero
in this limit, it has an almost degenerate horizon at zero radius, i.e.,
quasi-extreme black hole. In higher dimensions,
the derivative of the metric function $\bar{B}$ does not vanish at the horizon
and the solutions are non-degenerate.

In the context of the Lovelock gravity~\cite{Lovelock, Zumino}, the first,
second and third Lovelock actions are cosmological constant, Einstein-Hilbert
action, and the GB term, respectively. In four dimensions, the GB term is
known as Euler density and does not contribute to the field equations
at the classical level. In five and six dimensions, the GB term is the highest
order action  in the Lovelock gravity and would give significant deviation from
the four-dimensional case. For this reason and the qualitative
difference from the non-dilatonic case, we investigate the cases of
$D=4, 5, 6$ and $10$ in the following sections.

\section{$D=4$ black hole solutions}

We first present the black hole solutions for $D=4$.
Because there is ambiguity in the choice of $\c$ as discussed in \S 2,
we examine black hole solutions for the two different choices $\c=1$ and
$\c=\frac12$. The first one is examined in order to check
the consistency of our solutions with those in Ref.~\citen{TYM}.
The second case is examined to see how the results depend on the choice
of the dilaton coupling.

\subsection{$\c=1$ solutions}

The condition of the regular horizon~\p{phiprime} reduces to
\bea 
\tr_H^2\phi_H'^2-\frac{\tr_H\phi_H'}{C \gamma }+6=0,
\label{phip4}
\ena 
which can be shown to be equivalent to the condition given in
Ref.~\citen{TYM}.\footnote{
In this comparison, we should take into account the fact that the normalization
of $\tr$ differs by a factor $\sqrt{8}$ and the derivatives are also different.
}
Eq.~\p{phip4} has two solutions
\bea 
\phi_{H,\pm}' = \frac{1 \pm \sqrt{1 -24 C^2 \c^2}}{2C\c \tr_H},
\label{phip4_2}
\ena 
among which only the smaller solution gives regular black holes.\footnote
{In Ref.~\citen{TYM}, it is stated that the larger solution gives regular
solution, but this should be corrected.}
The solutions are real only for 
\bea 
\label{phi-del}
\tr_H^2\geq 4\sqrt{6}\gamma e^{-\frac{\gamma}{2}\phi_H},
\ena
which gives the lower bound on $\tr_H$ for the regular solution to exist
for any $\gamma$.
We will see that this is in sharp contrast to higher-dimensional solutions.

\begin{figure}
\includegraphics[width=8cm]{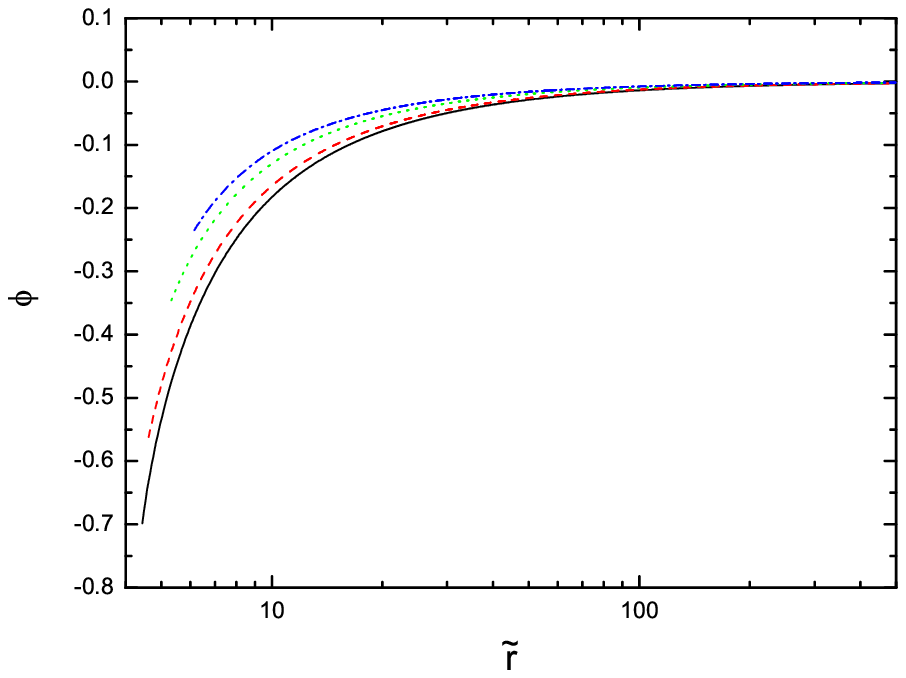}
\put(-112,-15){(a)}
\includegraphics[width=8cm]{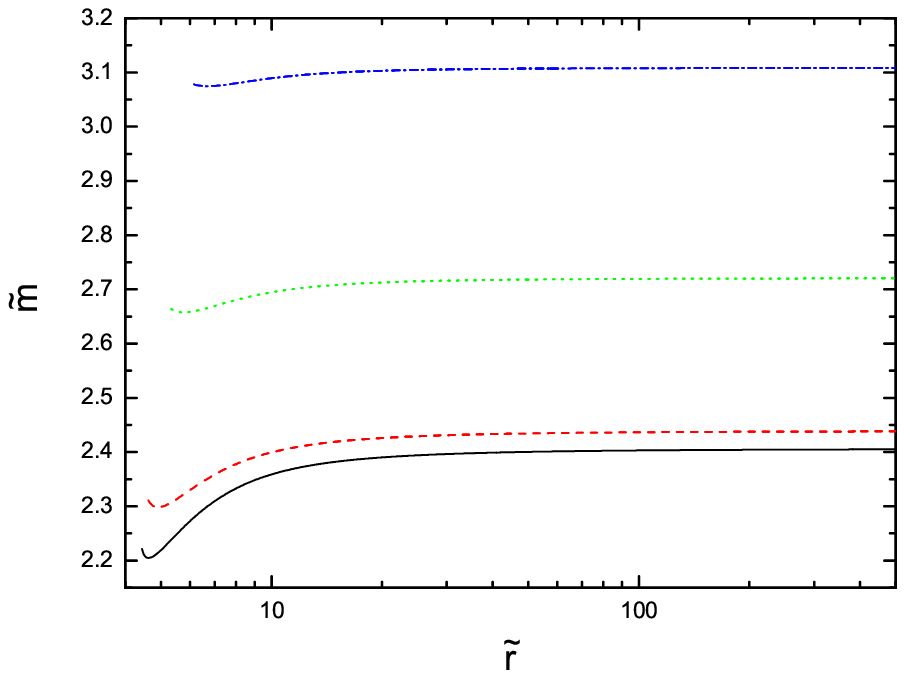}
\put(-110,-15){(b)}
\\
\vspace*{3mm}\\
\includegraphics[width=8cm]{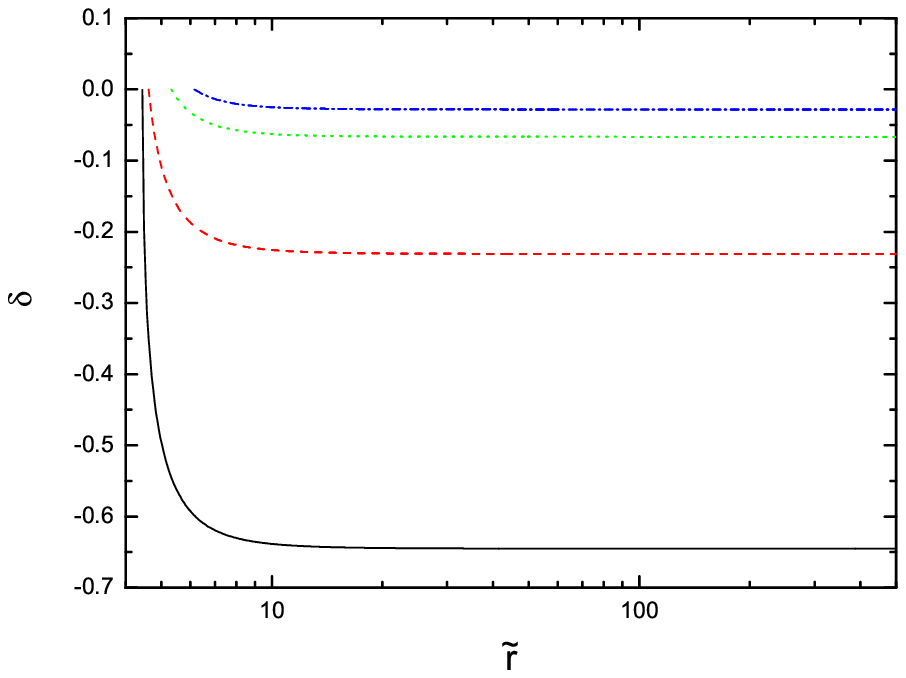}
\put(-112,-15){(c)}
\includegraphics[width=8cm]{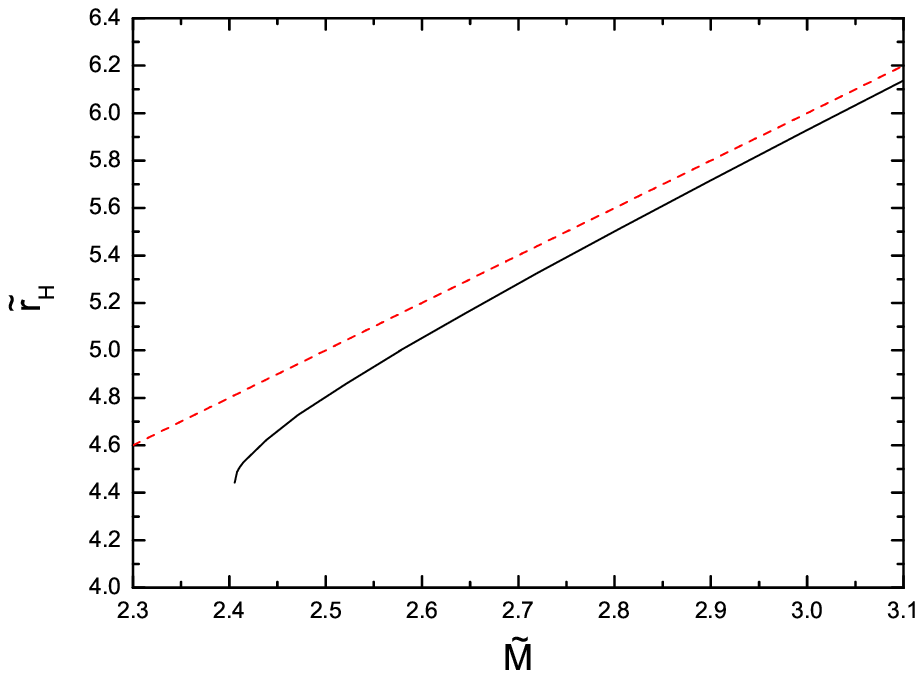}
\put(-110,-15){(d)}\\
\vspace*{3mm}\\
\includegraphics[width=8cm]{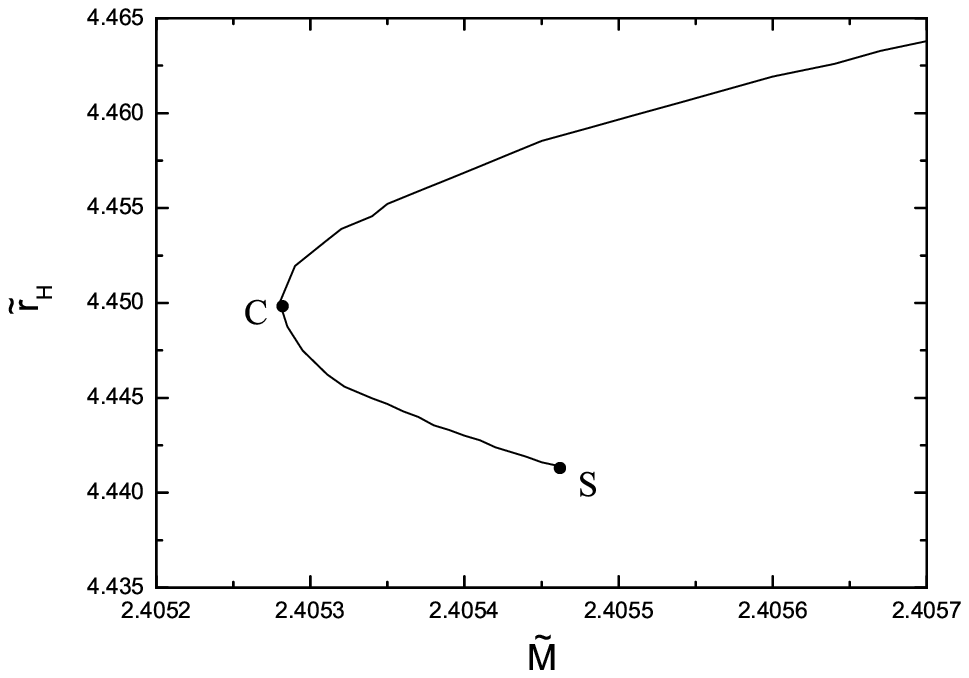}
\put(-112,-15){(e)}
\vspace*{3mm}\\
\caption{Black hole solutions in the four-dimensional Einstein-GB-dilaton
system with $\c=1$.
The behaviours of (a) the dilaton field, (b) the mass
function and (c) the lapse function for four different radii of event
horizon: $\tr_H=r_H/\sqrt{\a_2}=4.44251$ (solid line), $4.6226$
(dashed line), $5.32723$ (dotted line) and $6.15605$ (dash-dotted line).
(d) The mass versus horizon radius. The solid line is for the dilatonic case
while the dashed line is for the non-dilatonic case.
(e) The magnification of (d) around the turning point C (2.40528, 4.4500)
and the singular point S  (2.40546, 4.44142).
}
\label{d4}
\end{figure}

For several boundary conditions on $\phi_H$ and $\d_H$ at the horizon
with the smaller solution for $\phi'_H$ in \p{phip4_2}, we obtain the behaviors
of the dilaton, the mass and the lapse functions by integrating the basic
equations~\p{g1}--\p{g3} from the horizon.
Using the symmetry~\p{shift}, we set the asymptotic value of the dilaton to zero.
The resulting configurations of these functions are
depicted in Fig.~\ref{d4} for $\tr_H=4.44251$, $4.6226$, $5.32723$ and $6.15605$.
The masses $\tilde M$ for these cases are found to be 2.40541, 2.43848, 2.72056
and 3.10867, respectively.
We find that the regular black hole solutions exist only for $\tr_H \geq 4.44142$,
which is a consequence from the condition~\p{phi-del}.

In the present four-dimensional case, we find that
the dilaton field increases monotonously from the value at the horizon to zero
(Fig.~\ref{d4}(a)).
The mass function decreases near the horizon and increases towards
a finite value (Fig.~\ref{d4}(b)) as $\tr$ increases. In fact, it can be shown that
\bea 
\label{mp4}
\tilde{m}_H' = -\frac{1}{12} \tr_H^2\phi_H^{\prime 2},
\ena 
so the decrease near the horizon occurs for any solution.
If we regard the dilaton terms and those proportional to $\alpha_2$ in Eq.~(\ref{GB-eq})
as ``matter terms", the effective energy density is given by
$\rho_{\rm eff}=-T^t_t=  (D-2)\tm'/8\pi\tr^{D-2}$.
In the region where the mass function decreases, we see that
the effective energy density becomes negative.
This does not mean that these solutions are unstable, but they are
indeed stable.~\cite{TYM,torii}

A notable feature in these solutions is that there are two solutions with
different horizon radii for the range $2.40528 \leq \tilde M \leq 2.40546$
(Fig.~\ref{d4}(e)). In the non-dilatonic case, we have
\begin{equation} 
\tilde{M}=\frac12 \tilde{r}_H,
\end{equation} 
which is also displayed in Fig.~\ref{d4}(d)
and there can be a black hole however small the mass is.
But in the dilatonic case, we do not find analogous behavior.
For the solution with smallest horizon radius, the condition \p{phi-del}
is saturated, and the second derivative of the dilaton field diverges at the horizon.
Moreover, it was shown that the smaller black hole solutions
for $2.40528 \leq \tilde M \leq 2.40546$
are unstable although the larger black hole solutions are stable.~\cite{TYM,torii}

All these results are consistent with those in Ref.~\citen{TYM}.

\begin{figure}
\includegraphics[width=8cm]{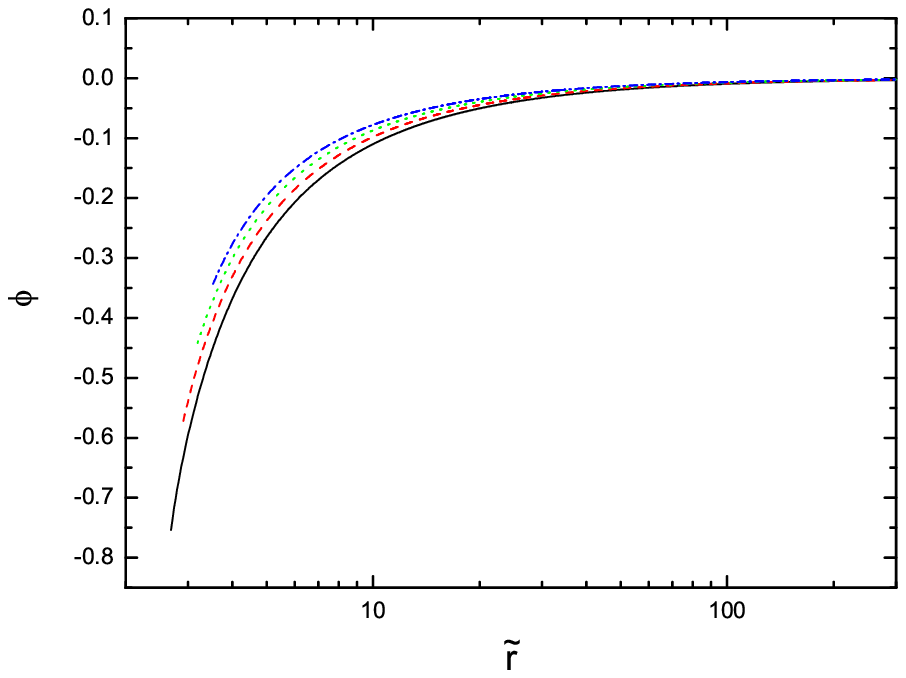}
\put(-112,-15){(a)}
\includegraphics[width=8cm]{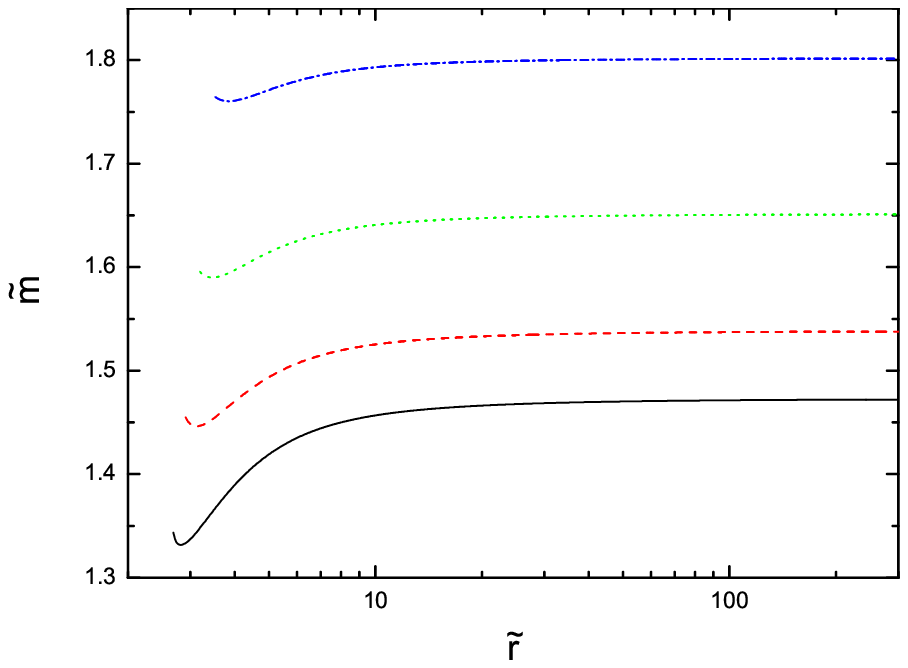}
\put(-110,-15){(b)}
\\
\vspace*{3mm}\\
\includegraphics[width=8cm]{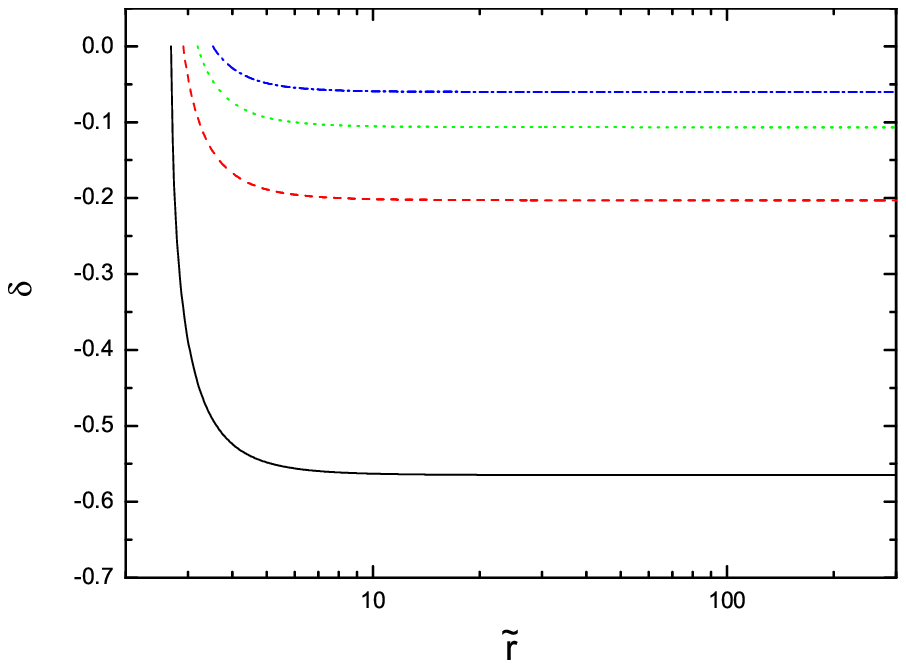}
\put(-112,-15){(c)}
\includegraphics[width=8cm]{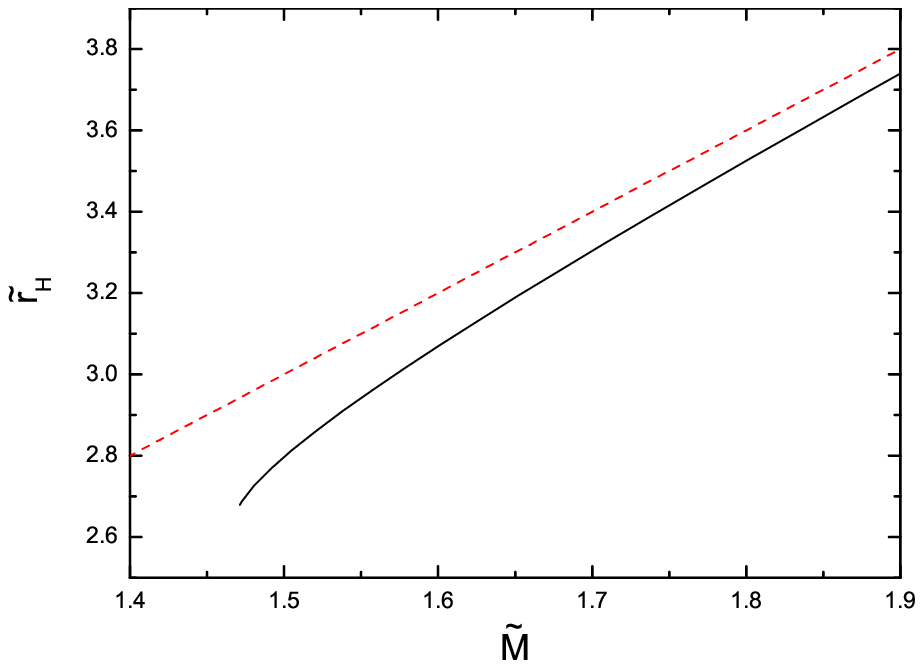}
\put(-110,-15){(d)}
\vspace*{3mm}
\caption{Black hole solutions in the four-dimensional Einstein-GB-dilaton
system with $\c=\frac12$. The behaviours of (a) the dilaton
field, (b) the mass function and (c) the lapse function for four
different radii of event horizon: $\tr_H=r_H/\sqrt{\a_2}=2.68697$
(solid line), $2.90965$ (dashed line), $3.19148$ (dotted line) and
$3.52851$ (dash-dotted line).
(d) The mass versus horizon radius in the dilatonic case (solid line)
and in the non-dilatonic case (dashed line).}
\label{d4-2}
\end{figure}

\subsection{$\c=\frac12$ solutions}

We next examine the $\c=\frac12$ case.

For several boundary conditions on $\phi_H$ and $\d_H$ at the horizon
with the smaller solution for $\phi'_H$ in \p{phip4_2}, we obtain the behaviors
of the dilaton, the mass and lapse functions by integrating the basic
equations~\p{g1}--\p{g3} from the horizon.
Using the symmetry~\p{shift}, we set the asymptotic value of the dilaton to zero.
The resulting configurations of these functions are
depicted in Fig.~\ref{d4-2} for $\tr_H=2.68697$, $2.90965$, $3.19148$ and $3.52851$.
The masses $\tilde M$ for these cases are found to be 1.47251, 1.53808, 1.65113,
and 1.80161, respectively.
We find that the regular black hole solutions exist only for $\tr_H \geq 1.47126$,
which is a consequence of the condition~\p{phi-del}.

We note that for this choice of $\c=\frac12$, there are not two solutions close
to the minimum value of the mass parameter in contrast to $\c=1$.
Except for this, we find that the behaviors of the dilaton, the mass and the
lapse functions are quite similar.
We expect that similar behaviors are obtained regardless of the value of $\gamma$
also in other dimensions.
So in the rest of this paper, we examine $\c=\frac12$ case only.

\section{$D=5$ solutions}

Five-dimension is the lowest dimension in which the GB term makes nontrivial
contribution to the vacuum solution. Eq.~\p{phiprime} reduces to
\begin{equation} 
C \gamma (1+C+3 C^2 \gamma ^2) \tr_H^2 \phi_H'^2
- (1+C) (1+C -6 C^2 \gamma^2) \tr_H \phi_H'
+\; 3 C (3-2 C-3 C^2) \gamma =0.
\label{phip5}
\end{equation} 
Here again, there are two solutions but only the smaller solution
of Eq.~\p{phip5} gives regular black holes.
The discriminant of this quadratic equation is (for $\c=\frac12$)
\bea
18 C^6+ 30 C^5+ 5 C^4- 16 C^3- 12 C^2+ 8 C +2,
\ena
which is always positive for $C>0$, and hence
there is no bound on the value of $\tr_H$ for the reality of the solution
like~\p{phi-del} in contrast to four dimensions.

\begin{figure}
\includegraphics[width=8cm]{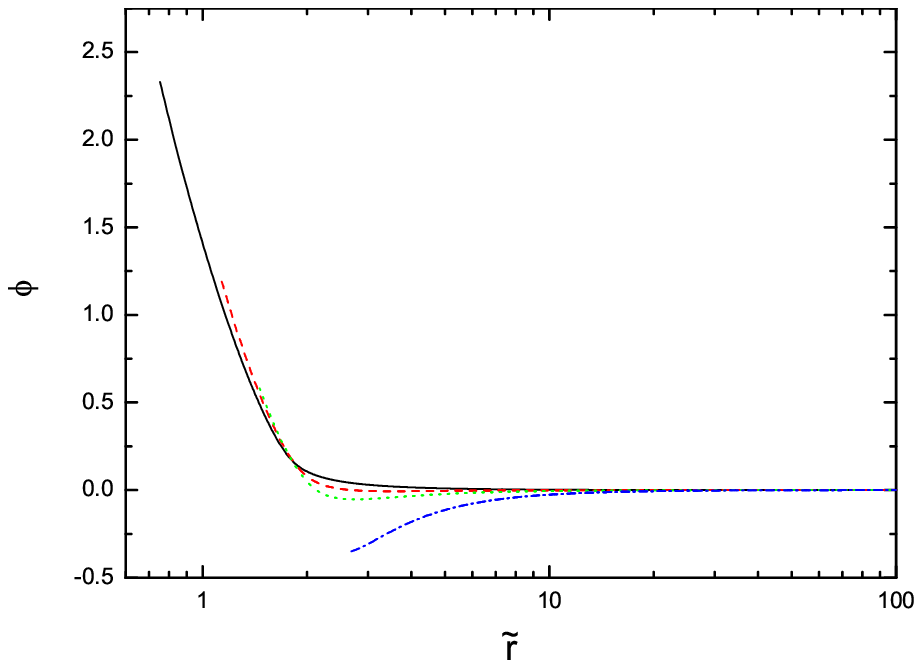}
\put(-112,-14){(a)}
\includegraphics[width=8cm]{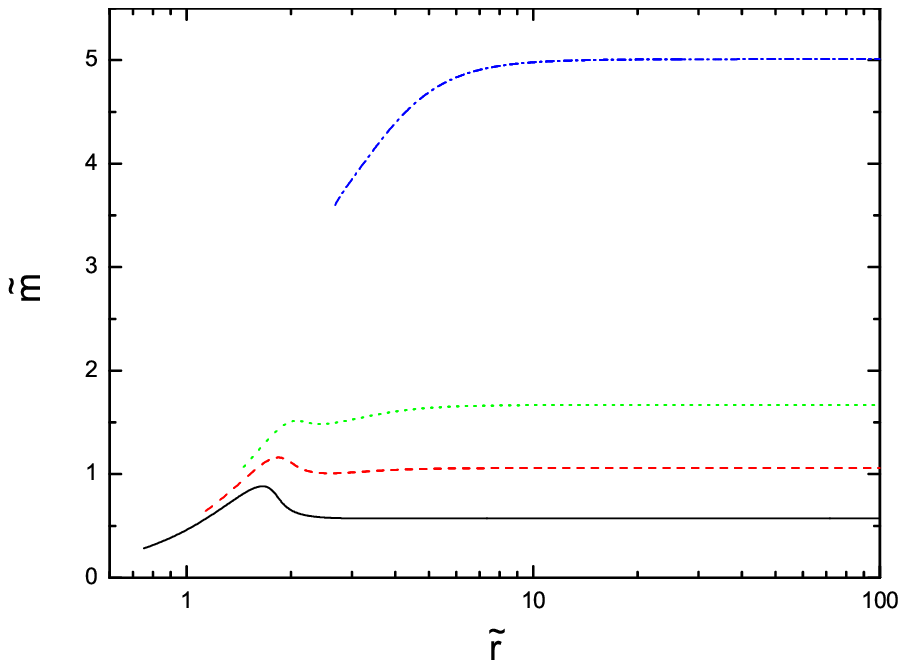}
\put(-110,-14){(b)}
\vspace*{1mm}\\
\includegraphics[width=8cm]{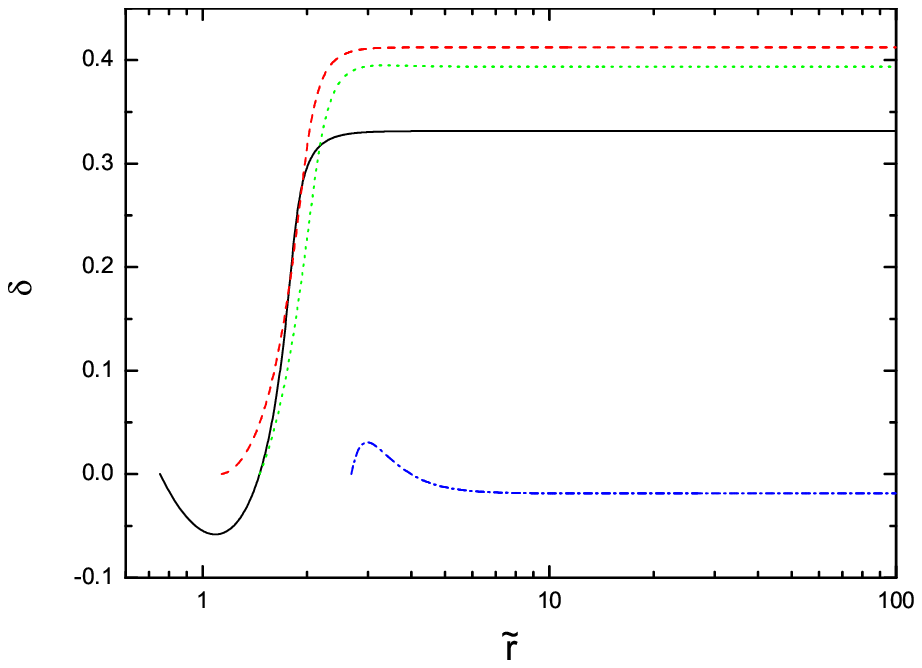}
\put(-112,-14){(c)}
\includegraphics[width=8cm]{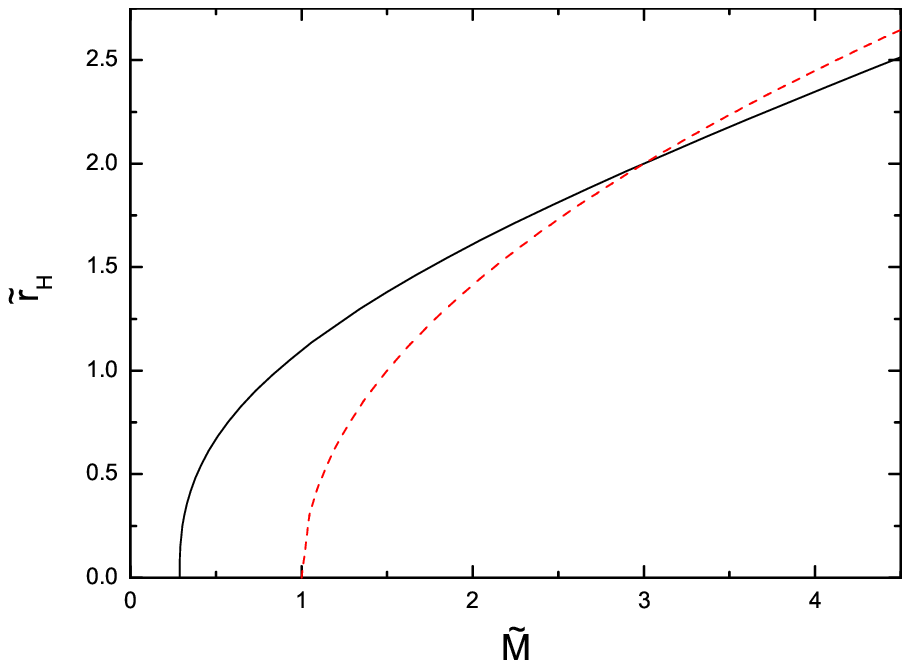}
\put(-110,-14){(d)}
\vspace*{1mm}\\
\includegraphics[width=8cm]{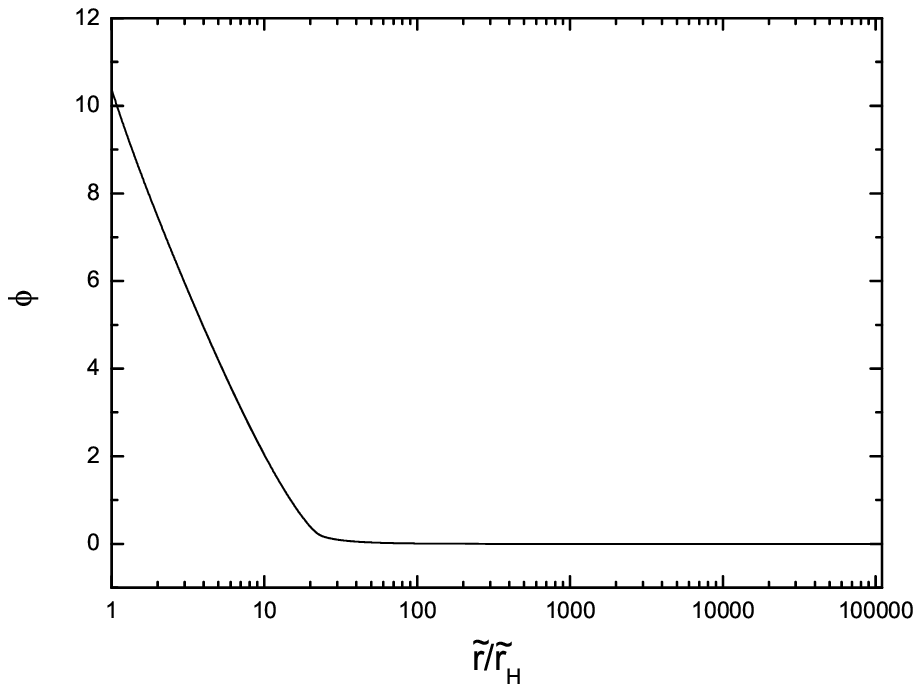}
\put(-112,-13){(e)}
\caption{Black hole solutions in the five-dimensional Einstein-GB-dilaton
system.
The behaviours of (a) the dilaton field, (b) the mass
function and (c) the lapse function for four different radii of
event horizon: $\tr_H=r_H/\sqrt{\a_2}=0.754129$ (solid line),
$1.13599$ (dashed line), $1.46193$ (dotted line) and $2.68391$ (dash-dotted line).
(d) The mass versus horizon radius for the dilatonic black holes (solid line) and
the non-dilatonic holes (dashed line).
The mass of the dilatonic black holes approaches a non-zero constant
$\tilde{M} = 0.288185$ as $\tr_H \to 0$.
(e) The configuration of the dilaton field for the small black holes with
$\tr_H=r_H/\sqrt{\a_2}=0.0748464$.
In the GB region the dilaton decays logarithmically and
suddenly changes to the power decay $\sim r^{-2}$.}
\label{d5}
\end{figure}

For various boundary conditions for $\phi_H$ and $\d_H$ and smaller solution
$\phi'_H$ of Eq.~\p{phip5} at the horizon, we obtain the configurations of
the dilaton field, the mass and lapse functions, which are depicted
in Fig.~\ref{d5} for $\tr_H=0.754129, 1.13599, 1.46193$ and $2.68391$.
The masses $\tilde M$ for these cases are found to be $0.573328$, $1.05972$,
$1.66924$ and $5.0097$, respectively.
In contrast to the four-dimensional case, we find that the regular black hole
solutions exist for all $\tr_H > 0$, in accordance with the above observation.

For the large black holes, the dilaton field $\phi$ monotonously increases
just as in the four-dimensional case. (We note that the smallest black hole
in four dimensions is the solution at the turning point C in Fig.~\ref{d4},
which has $\tilde M=2.40528$). From Eq.~(\ref{phip5}),
we find that the smaller solution for $\phi'_H$ is positive when $C(>0)$
satisfies both the conditions
\begin{equation} 
C<\dfrac{\sqrt{10}-1}{3}
 ~~~ {\rm and} ~~~
C<\dfrac{1+\sqrt{1+24\gamma^2}}{12\gamma^2},
\end{equation} 
and is non-positive otherwise.
For $\gamma=\frac12$, the first condition is stronger than
the second one, and the condition becomes $C\lesssim 0.72$.
Since $\phi_H \to 0$ and $C\to 0$ for the large black hole, the dilaton field
increases at the horizon.
For the small black holes, however, the GB term
affects the configurations of the field functions significantly near the horizon,
and the dilaton field takes positive value at the horizon. Then
it decreases asymptotically. This means that the dilaton charge is negative.

For the black holes with large mass, the mass function increases monotonously,
as in the four-dimensional case. It
shows, however, peculiar behaviors for small black holes.
It increases near the horizon but decreases in the intermediate region,
which means that the effective energy density $\rho_{\rm eff}\sim \tm'/\tr^{D-2}$
becomes negative there. These can be intuitively explained as follows.
There are two typical length scales for our solution. One is the string scale
given by $\ell_s=\sqrt{\alpha_2}$
\footnote{This differs from the usual definition of the string length by a factor
$\sqrt{8}$, but we are interested only in the order of magnitude.}
and the other is the horizon radius $r_H$.
When the length scale we are interested in is longer than $\ell_s$, the GB term
is negligible in the field equations (\ref{GB-eq}) and (\ref{dil-eq}).
On the other hand, when the length scale is shorter than $\ell_s$, the effect
of GB term becomes dominant. Hence for the small black holes, the field functions
show qualitatively exotic behaviors in the region $r_H< r<\ell_s$.
In fact, the log-linear plots of the dilaton field (Fig.~\ref{d5} (e))
shows that the dilaton field behaves as $\sim \log \tr$ in this region. Beyond this
region, the behavior suddenly changes to power decay $\sim \Sigma/\tr^2$.
This is a characteristic feature of the solution in which spacetime is divided
into the GB region and the GR region distinctly.

The fact that the effective energy density is negative in some intermediate
region does not mean that the solutions are unstable. In fact, it has been
checked that they are stable in four dimensions. We expect the same stability here.

The perturbative approach gives us clearer and further understanding.
Let us assume $\gamma\ll 1$ and expand the basic equations (\ref{GB-eq})
and (\ref{dil-eq}) with respect to $\gamma$. Then the zeroth order equations
reduce to the Einstein-GB-massless scalar system and give the non-dilatonic solution
with $\phi\equiv 0$. Since the terms with the dilaton field in the gravitational
equation (\ref{GB-eq}) are second order, the non-dilatonic solution is also
the solution of Eq.~(\ref{GB-eq}) up to the first order.

On the other hand, the first order equation for the dilaton field gives
\begin{equation} 
\bar{\squaret} \phi_1 = \alpha_2  \bar{R}_{\rm GB}^2,
\label{per-phi0}
\end{equation} 
where $\phi_1$ is the first order expansion of $\phi$ as
$\phi=\phi_0+\phi_1 \gamma +\cdots$, and $\bar{\squaret}$ and $\bar{R}_{\rm GB}^2$
are evaluated by the zeroth order variables.
Under our ansatz, Eq.~(\ref{per-phi0}) is written as
\begin{equation} 
\frac{1}{\tr^{D-2}}\bigl(\tr^{D-2} \phi_1'\bigr)'\bar{B}+\bar{B}'\phi_1'
=  \bar{R}_{\rm GB}^2.
\end{equation} 
(It should be noted that the variables are normalized by $\sqrt{\alpha_2}$.)
In the asymptotic region, the first term in the left had side is the
leading term and we find $\phi_1 \sim -\Sigma/\tr^{D-3}$. We call it the GR region.
In the intermediate region, this equation can be integrated as
\begin{equation} 
\label{per-phi}
\phi_1(\tr) = \int^{\tr}_{\tr_H} \frac{1}{\tr^{D-2}\bar{B}}\biggl[
\int^{\tr}_{\tr_H} \tr ^{D-2}\bar{R}_{\rm GB}^2 d\tr \biggr]d\tr +
\phi_H.
\end{equation} 
Near the event horizon, the behavior of the dilaton $\phi$ is nontrivial,
and this is due to the effect of the GB term. We call it the GB region.

To see the boundary of between the GR region and the GB region,
the discussion of the  the effective energy density is helpful.
Let us investigate the effective energy density.
In the perturbative approach, we find
\begin{equation} 
\label{eff-ene}
\rho_{\rm eff}=\frac{(D-1)_2}{32\pi(D-3)_4}
\biggl(\frac{1+(\mu/\tr)^{D-1}}{\sqrt{1+2(\mu/\tr)^{D-1}}}-1\biggr),
\end{equation} 
where
\begin{equation} 
\label{mu}
\mu=\Bigl\{2(D-3)_4\tr_H^{D-5}\bigl[\tr_H^{2}+(D+3)_4\bigr]\Bigr\}^{\frac{1}{D-1}},
\end{equation} 
and we have chosen the GR branch.
When the condition $\tr \gg \mu$ is satisfied,
\begin{equation} 
\rho_{\rm eff}=\frac{(D-1)_2}{64\pi(D-3)_4}\Bigl(\frac{\mu}{\tr} \Bigr)^{2(D-1)} \sim 0.
\end{equation} 
Since the effective energy density vanishes, this condition corresponds to the GR region.
When the condition $\tr \ll \mu$ is satisfied,
\begin{equation} 
\rho_{\rm eff}
=\frac{(D-1)_2}{32\pi(D-3)_4}\sqrt{\frac12\Bigl(\frac{\mu}{\tr}\Bigr)^{D-1}}.
\end{equation} 
Hence this region is significantly affected by the GB term, and it is the GB region.
The boundary $\tr_{\ast}$ of the transition between the GR region and the GB
region is simply obtained by $\tr_{\ast} = \mu$. In the five-dimensional case,
this condition becomes
\begin{equation} 
\frac{r_{\ast}}{\ell_s}
=\biggl[4\Bigl(\frac{r_H^2}{{\ell_s}^2}+2\Bigr)\biggr]^{\frac{1}{4}}.
\end{equation} 

For the large black holes with $r_H \gg \ell_s$, the boundary is
at $r_{\ast} =\sqrt{2r_H \ell_s}$. This is less than $r_H$ and the transition
occurs inside the event horizon. Hence the outer spacetime of the event horizon
is well approximated by Tangherlini solution, ($D$-dimensional Schwarzschild
solution)~\cite{Tangherlini}. For the ``string size" black holes with $r_H = \ell_s$,
the transition occurs around $r_{\ast} \sim \sqrt{2} \ell_s$. The effect of GB term
extends to the region of a few times of radius of event horizon.

For the small black holes with $r_H \ll \ell_s$, the boundary is $r_{\ast} \sim \ell_s$.
Hence the string effect extends to the region of the string scale.
These results are consistent with the intuitive consideration given above.

We have found that the regular black hole solutions exist for all $\tr_H > 0$.
In the four-dimensional case, $\phi'_{H,\pm}$ 
Eq.~(\ref{phip4}) degenerate for the finite horizon radius, and the solution
disappears below the radius~\p{phi-del}.
On the other hand, $\phi'_H$ has two non-degenerate roots even for the $\tr_H \to 0$
limit in five dimensions, and there are regular black hole solutions
for all $\tr_H$.
The mass of the black hole approaches a nonzero constant $\tilde{M} = 0.288185$
as $\tr_H \to 0$.

It is instructive to compare this result with the non-dilatonic case.
For five dimensions, Eq.~\p{nondil} gives
\begin{equation} 
\bar{M}=\frac12 (\tilde{r}_H^2 +2).
\end{equation} 
As $\tilde{r}_H \to 0$, the mass of the black hole approaches
a nonzero constant $\bar{M} \to 1$.
This property is very similar to our dilatonic case although the constant
value is different due to the effect of the dilaton.

The difference of masses between the dilatonic and non-dilatonic cases
(Fig.~\ref{d5}(d)) can be also estimated in terms of the length scales.
For the large black holes ($r_H>\ell_s$), the GB term can be neglected and the
dilaton field contributes to the solutions mainly as an ordinary matter field
in GR, and the mass becomes large compared to the non-dilatonic case. However,
for the small black holes ($r_H<\ell_s$), the effect of the GB term becomes large.
If we regard the GB term as a matter term again, the dilaton field plays the role of
the coupling between the GB term and the gravity. Because of this nontrivial
coupling with the GB term, the effective energy density becomes negative
(see Fig.~\ref{d5}(b)), and the total mass decreases compared to the non-dilatonic
case. In the zero horizon-radius limit, $C$ and $\tr_H \phi'_H$ take constant values,
which means that the derivative of the dilaton field diverges at the horizon.
The dilaton field itself also diverges as $\phi_H \sim -2\log \tilde{r}_H/\gamma$.
As a result, the energy density of the dilaton field diverges.
Since the derivative of the mass function also diverges at the horizon,
the zero size black hole is singular.
The difference between $\tilde M$ in the dilatonic case and $\bar{M}$ in
the non-dilatonic case in the $\tr_H\to 0$ limit is due to the diverging
energy density of the dilaton field at $\tr_H =0$.

\section{$D=6$ solutions}

We now examine the six-dimensional solutions. Actually we find that many
of the properties in the $D= 6$ case are similar to the higher dimensional cases
up to ten dimensions, where we come back to the original string theory.
So we will be very brief, leaving more detailed discussions to $D=10$.

For $D=6$, Eq.~\p{phiprime} reduces to
\bea 
&& C \gamma \Big[6+14 C+4 C^2 (14 \gamma ^2+1)+48 C^3 \gamma ^2\Big] \tr_H^2\phi_H'^2
\nn
&&
\hs{5}+2\Big[40 C^2\gamma ^2 (1+C -C^2)-(3 +C) (1+2 C)^2\Bigr]\tr_H \phi_H'
+ 40 C (2 -2 C-7 C^2) \gamma =0.~~
\label{phip6}
\ena 
Here again, there are two solutions but only the smaller solution
of Eq.~\p{phip6} gives regular black holes.
The discriminant of this quadratic equation is (for $\c=\frac12$)
\bea
100 C^8 + 720 C^7 + 1656 C^6 + 1288 C^5 + 280 C^4 - 40 C^3 + 85 C^2 + 78 C +9,
\ena
which is again always positive for $C>0$, and hence
there is no bound on the value of $\tr_H$ for the reality of the solution.

\begin{figure}
\includegraphics[width=8cm]{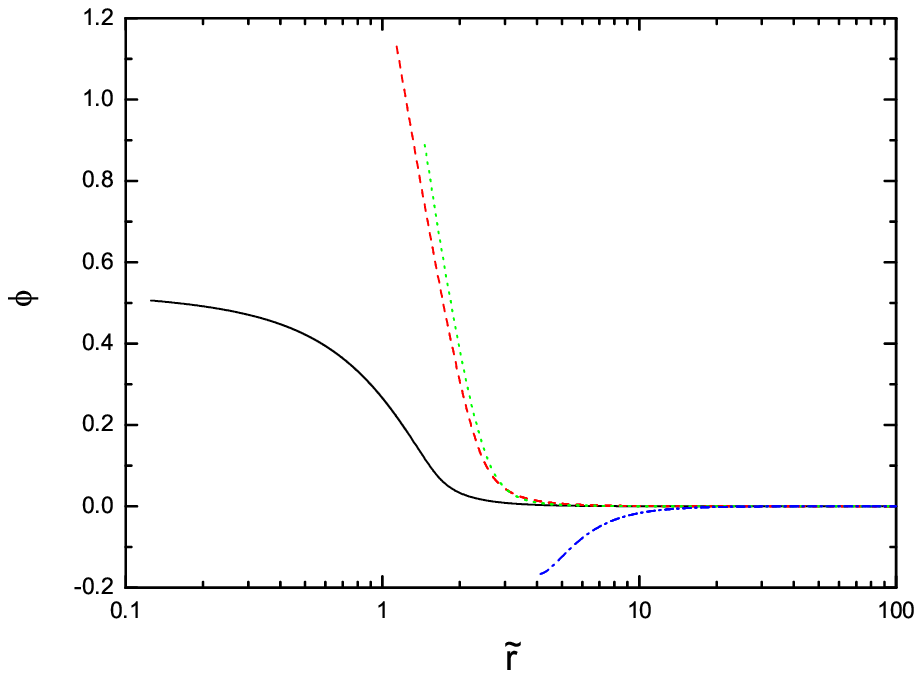}
\put(-112,-14){(a)}
\includegraphics[width=8cm]{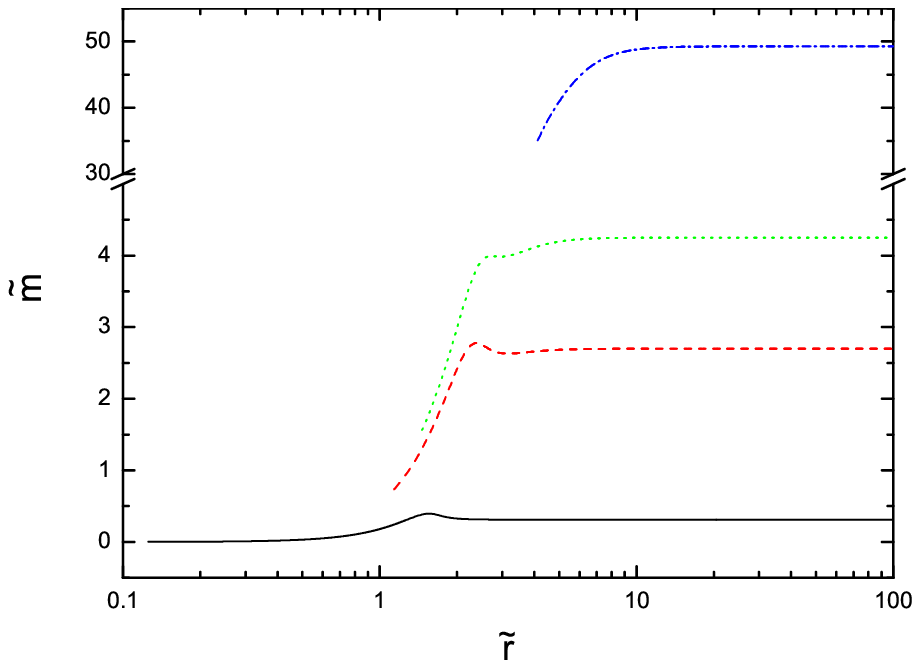}
\put(-110,-14){(b)}
\vspace*{1mm}\\
\includegraphics[width=8cm]{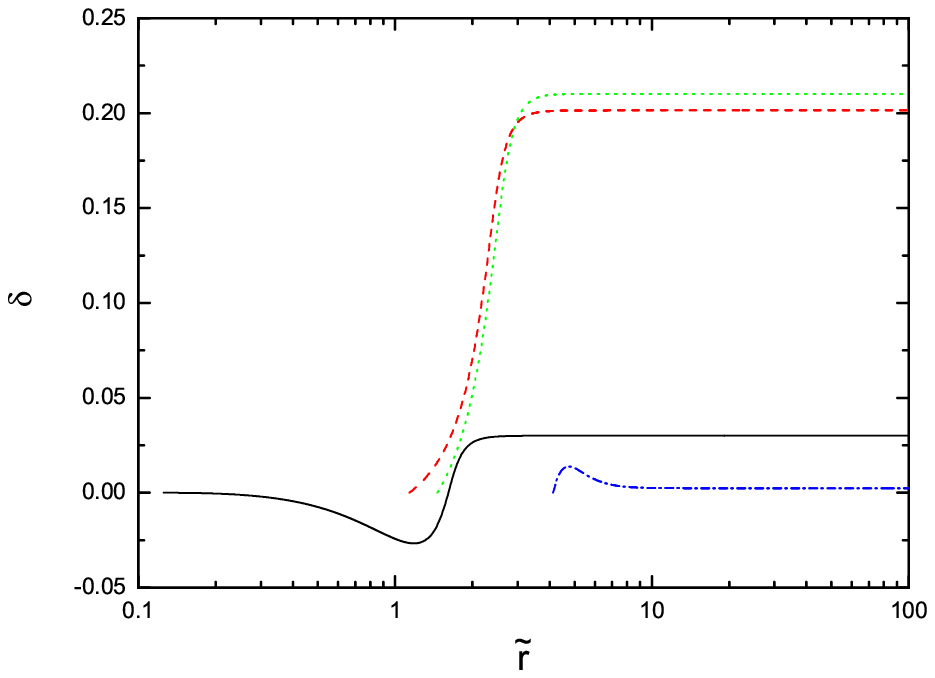}
\put(-112,-14){(c)}
\includegraphics[width=8cm]{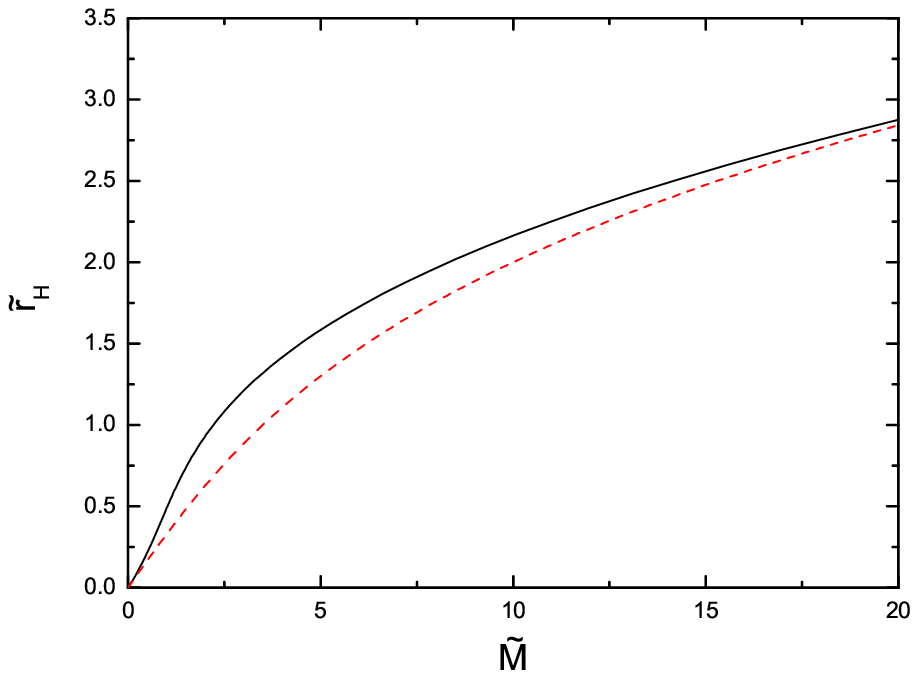}
\put(-110,-14){(d)}
\caption{Black hole solutions in the six-dimensional Einstein-GB-dilaton
system. The behaviour of (a) the dilaton field, (b) the mass function
and (c) the lapse function for four different radii of the event horizon:
$\tr_H=r_H/\sqrt{\a_2}=0.125367$ (solid line), $1.13596$ (dashed
line), $1.46199$ (dotted line) and $4.12369$ (dash-dotted line).
(d) The mass versus horizon radius for the dilatonic black holes (solid line) and
the non-dilatonic holes (dashed line).
The mass of the black hole approaches $0$ as $\tr_H \to 0$.}
\label{d6}
\end{figure}

For various boundary conditions for $\phi_H$ and $\d_H$ and smaller solution
$\phi'_H$ of Eq.~\p{phip6} at the horizon, we find the behaviors of the dilaton,
the mass and lapse functions, which are depicted in Fig.~\ref{d6}
for $\tr_H=0.125367$, $1.13596$, $1.46199$ and $4.12369$.
The masses $\tilde M$ for these cases are found to be $0.311672$, $2.6993$,
$4.25081$ and $49.2744$, respectively.

For the large black holes, the dilaton field $\phi$ monotonously increases as
in the four-dimensional case, and for the smaller black holes ($r_H\sim \ell_s$),
$\phi_H$ becomes large and the dilaton field decreases asymptotically with
the negative dilaton charge as in the five dimensional case.
However, as the black hole becomes further smaller ($r_H\ll \ell_s$), the
magnitude of the dilaton field becomes smaller and vanishes in the $r_H \to 0$ limit.

As for the mass function $\tm$, it has a peak and there is a region where
the effective energy density is negative when the horizon radius is $r_H\sim \ell_{s}$.
This means that the string effect is significant. However, as the black hole
becomes smaller, the mass function have no peak and takes small value,
which vanishes in the zero horizon-radius limit.

We find that the regular black hole solutions exist for all $\tr_H > 0$.
The mass of the black hole approaches zero as $\tr_H \to 0$.
This is different from the four- and the five-dimensional cases,
but it is in agreement with the non-dilatonic case.
In fact, for six dimensions, Eq.~\p{nondil} gives
\begin{equation}
\bar{M}=\frac12 {\tr_H} (\tr_H^2 +6).
\end{equation}
As $\tilde{r}_H \to 0$, the mass of the black hole approaches zero.

We now turn to the discussions of the ten-dimensional solutions.

\section{$D=10$ solutions}

The $D=10$ is the most interesting case from the theoretical point of view
since it is the critical dimension of string theory. In the Einstein-Maxwell-dilaton
system, the spacetime structure of the $D$-dimensional black hole solutions
changes at ten dimensions~\cite{GM, GHS}.

For $D=10$, Eq.~\p{phiprime} reduces to
\bea 
&& C \gamma  \Bigl[7 +57 C +6 C^2 (76 \gamma ^2+15)
+1680 C^3 \gamma^2\Bigr] \tr_H^2\phi_H'^2
\nn
&&\hs{10}+\Bigl[432 C^2 \gamma ^2(1+C-15 C^2)  - (1+6 C)^2 (7+15 C) \Bigr]\tr_H\phi_H'
\nn
&&
\hs{10} -72 C (-4+6 C+99 C^2) \gamma=0.
\label{phip10}
\ena 
Once again, the discriminant of this quadratic equation is (for $\c=\frac12$) always
positive for $C>0$, and there is no bound on the value of $\tr_H$ for the reality
of the solution.

\begin{figure}
\includegraphics[width=8cm]{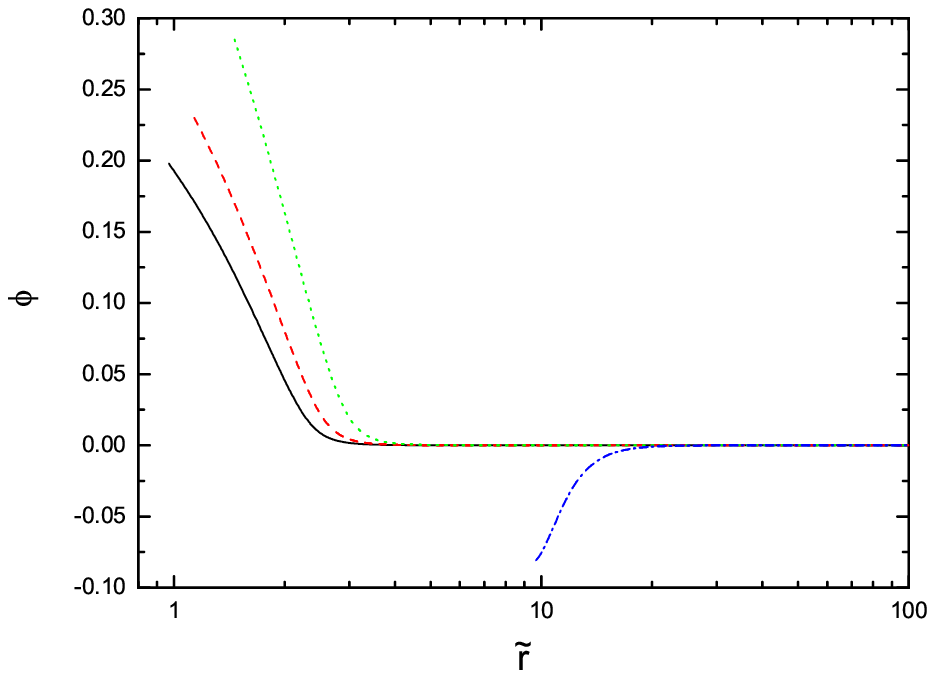}
\put(-112,-14){(a)}
\includegraphics[width=8cm]{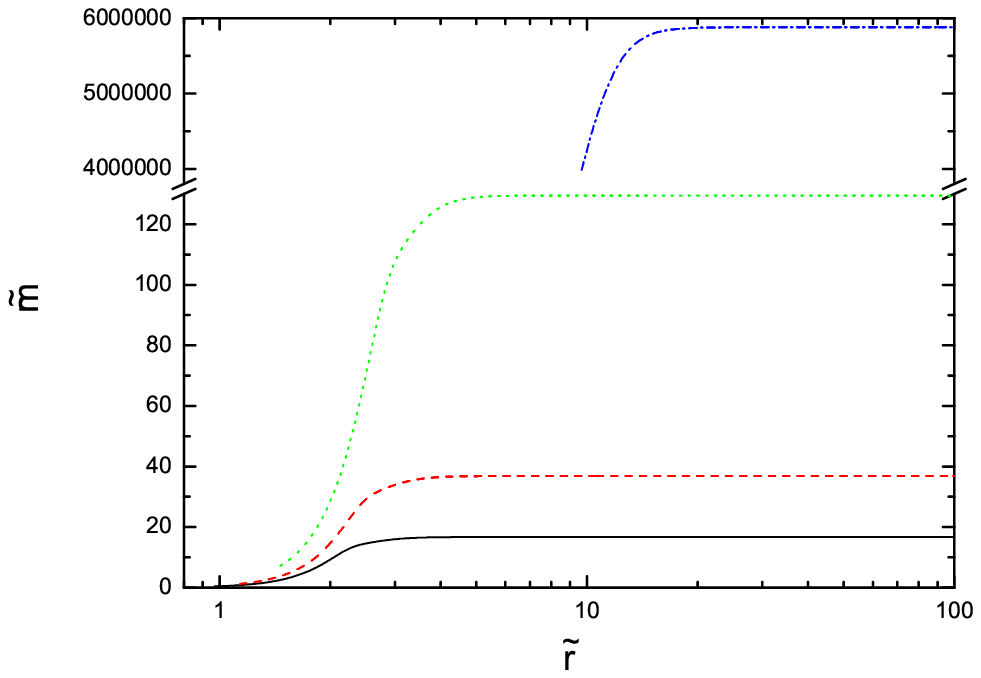}
\put(-110,-14){(b)}
\vspace*{1mm}\\
\includegraphics[width=8cm]{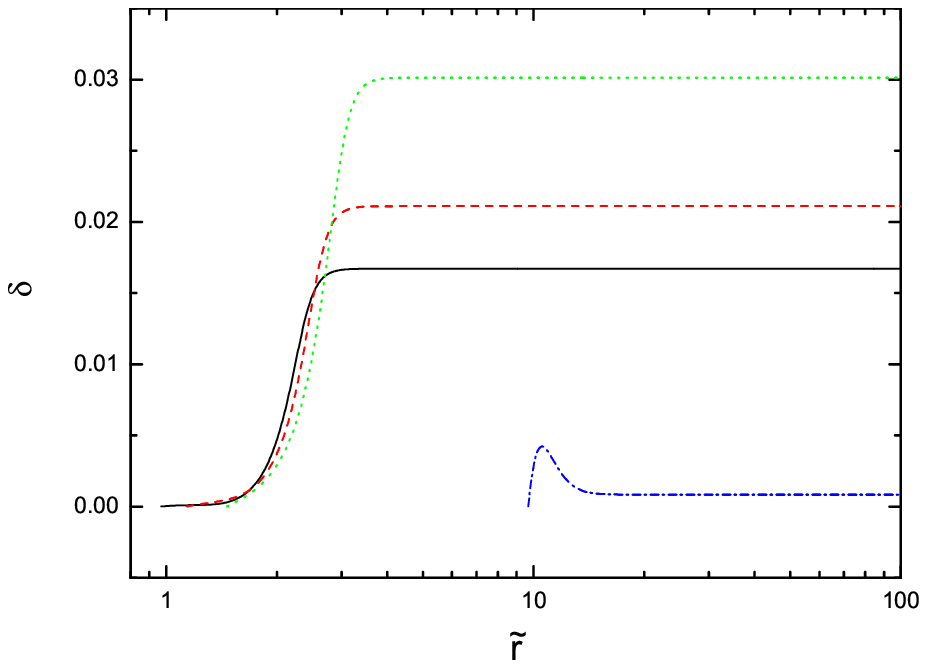}
\put(-112,-14){(c)}
\includegraphics[width=8cm]{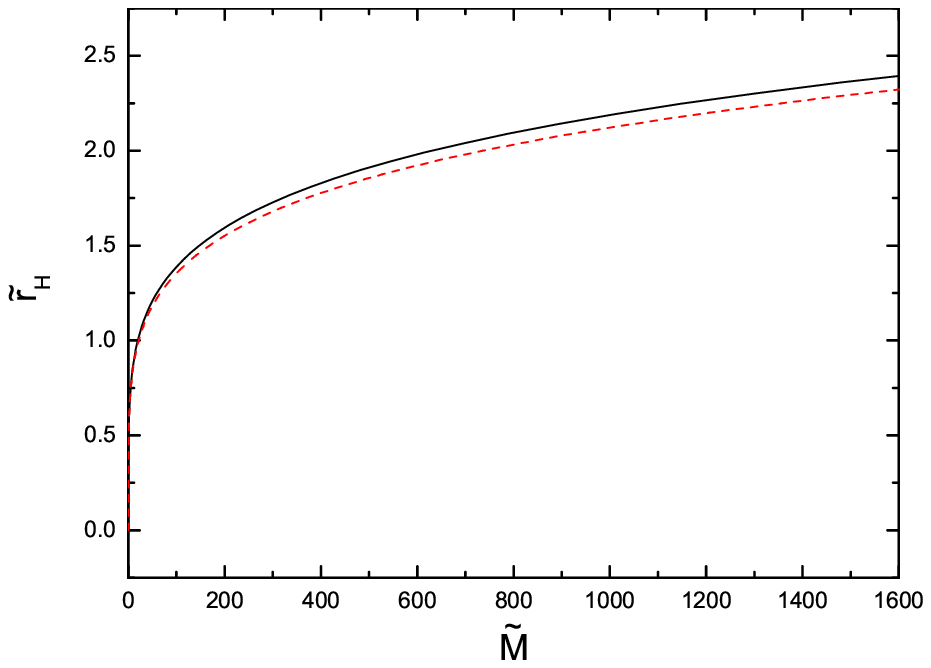}
\put(-110,-14){(d)}
\caption{Black hole solutions in the ten-dimensional Einstein-GB-dilaton
system. The behaviour of (a) the dilaton field, (b) the mass function
and (c) the lapse function for four different radii of the event horizon:
$\tr_H=r_H/\sqrt{\a_2}=0.968549$ (solid line), $1.13596$ (dashed line),
$1.46194$ (dotted line) and $9.68119$ (dash-dotted line).
(d) The mass versus horizon radius for the dilatonic black holes (solid line) and
the non-dilatonic holes (dashed line).
The mass of the black hole approaches $0$ as $\tr_H \to 0$.}
\label{d10}
\end{figure}

For various boundary conditions for $\phi_H$ and $\d_H$ and smaller solution
$\phi'_H$ of Eq.~\p{phip10} at the horizon, we find the behaviors of the dilaton,
the mass and the lapse functions, which are depicted in Fig.~\ref{d10}
for $\tr_H=0.968549$, $1.13596$, $1.46194$ and $9.68119$.
The masses $\tilde M$ for these cases are found to be $16.7172$, $36.8633$,
$129.489$ and $5.88035\times 10^6$, respectively.

The configurations of the field functions are almost the same as those in the $D=6$ case
qualitatively.
For the large black holes the dilaton field $\phi$ monotonically increases. From
Eq.~(\ref{phiprime}), we find that the smaller solution for $\phi'_H$ is positive
when both the conditions
\begin{equation} 
C <  \frac{\sqrt{2D(D-1)}-2}{(D-4)(D+1)},
\end{equation} 
\begin{equation} 
(D-1)_2(D-4) C^2\bigl[2+2C-(D-4)_5 C^2\bigr] \gamma^2
-\bigl[1+(D-4)C\bigr]^2\bigl[2(D-3)+(D-4)_5 C\bigr]<0,
\end{equation} 
are satisfied, and  is non-positive otherwise.
Since $C\to 0$ in the large $\tr_H$ limit, both the conditions are satisfied
for the large black holes, which means that the dilaton field increases around
the horizon. This behavior is independent of the value of $\gamma$ and dimensions.
The difference from the $D=6$ case is that we find no peak in the mass function $\tm$,
and there is no region with negative effective energy density even for
the small black hole. This is in sharp contrast to other lower dimensions, and
may be related to the fact that the ten-dimension is the dimension that
the string theory lives in.
As $\tilde{r}_H \to 0$, the mass of the black hole
approaches zero. This is again in agreement with the non-dilatonic case,
for which Eq.~\p{nondil} gives
\begin{equation} 
\bar{M}=\frac12 \tr_H^5 (\tr_H^2 +42).
\end{equation} 

The behaviors of the functions are also explained in terms of the effective
energy density. In $D\geq 6$ case, $\mu$ is given by Eq.~(\ref{mu}),
and the boundary between the GR region and the GB region is
\begin{equation} 
\frac{r_{\ast}}{\ell_s}=\biggl\{2(D-3)_4\Bigl(\frac{r_H}{\ell_s}\Bigr)^{D-5}
\biggl[\Bigl(\frac{r_H}{\ell_s}\Bigr)^{2}+(D-3)_4\biggr]
\biggr\}^{\frac{1}{D-1}}.
\end{equation} 

For the large black holes with $r_H \gg \ell_s$, the boundary is
$r_{\ast} =[2(D-3)_4r_H^{D-3} \ell_s^2]^{\frac{1}{D-1}}$. This is less than
$r_H$ and the transition occurs inside the event horizon. Hence the outer
spacetime of the event horizon is the GR region.

For the ``string size" black holes with $r_H \simeq \ell_s$, the transition occurs
around $r_{\ast} \sim  \ell_s$. (The factor is $\approx 2.5$ for $D=10$.)
The effect of the GB term extends to the region of a few times of radius of
the event horizon. These are similar to the five-dimensional case.

For the small black holes with $r_H \ll \ell_s$, however, the boundary is
$r_{\ast} \ll \ell_s$. Hence the  string effect is confined to the region
just around the event horizon which is much smaller than the string scale.
One would naturally expect that the string effect extends to the region of
the string scale $\ell_s$. However, this is not the case.
Actually if we assume that the dilaton field does not diverge in the $r_H \to 0$
limit for $D\geq 6$, from Eq.~(\ref{phiprime}), we find
\begin{equation} 
\phi'_H=-\frac{D+1}{4(D-3)(D-5)\gamma}e^{\c\phi_H} \tr_H.
\end{equation} 
For $D=10$, this gives
\begin{equation} 
\phi'_H=-0.15714 \,e^{\c\phi_H}\tr_H,
\end{equation} 
which is consistent with what is obtained by numerical calculation where
$\phi_H =0$.
As a result, the dilaton field vanishes through the whole spacetime
and solution approaches the non-dilatonic one in the $\tr_H \to 0$ limit.
This behavior can be seen more clearly in six dimensions in
Fig.~\ref{d6}(d). The $\tilde{M}$-$\tr_H$ curve rapidly approaches
the non-dilatonic one in  the $\tr_H \to 0$ limit.

\section{Thermodynamics}

Thermodynamic property of black holes is one of the important issues,
in particular when we discuss the evolution of the black holes through Hawking
radiation. It is expected that the string effects become dominant
in the final stage of the black hole evaporation and give non-trivial processes.
Although black hole thermodynamics in non-GR theories is not well understood
in comparison with that in GR, we can define the temperature and entropy
of the black hole which obey the first law of thermodynamics in asymptotically
flat spacetime in diffeomorphism invariant theories~\cite{Wald}.
The Hawking temperature is given by the periodicity of the Euclidean time
on the horizon by
\bea 
\label{temp}
T_H =\frac{e^{-\d_H}}{4\pi r_H} \biggl(D-3-\frac{2 \tm_H'}{\tr_H^{D-4}}\biggr) .
\ena 

\begin{figure}
\includegraphics[width=8cm]{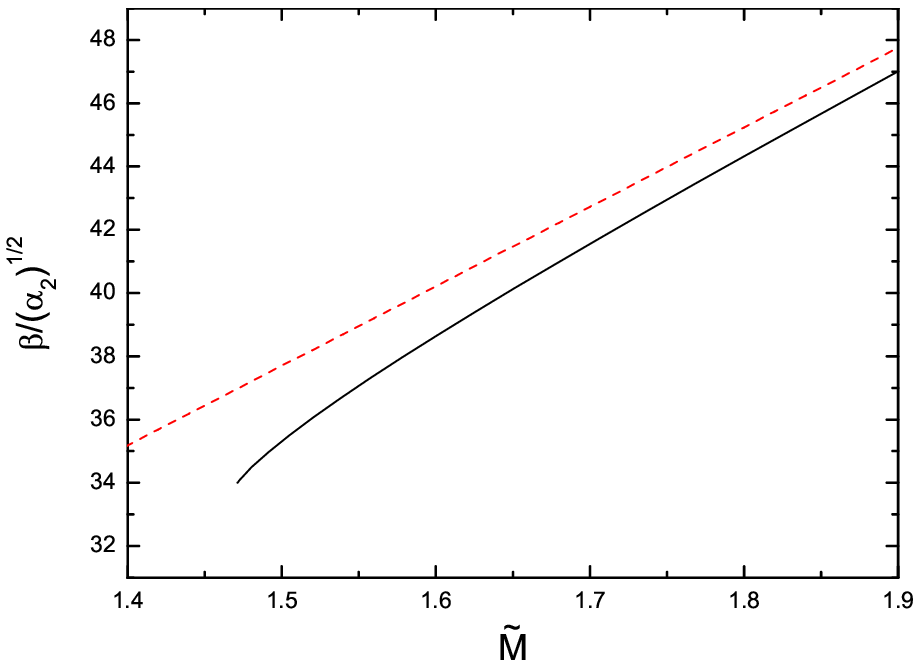}
\put(-112,-15){(a)}
\includegraphics[width=8cm]{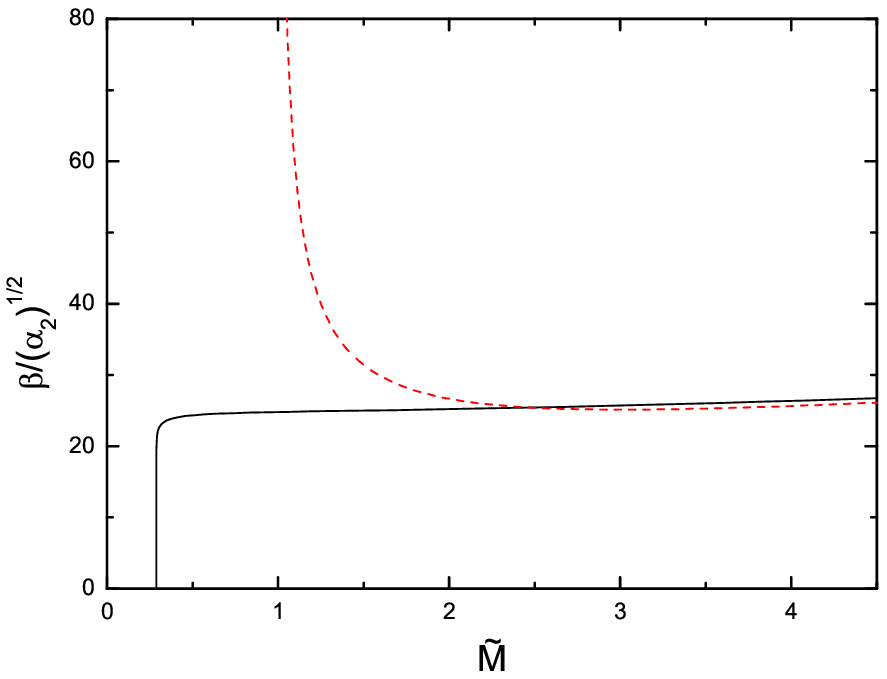}
\put(-110,-15){(b)}
\vspace*{1mm}\\
\includegraphics[width=8cm]{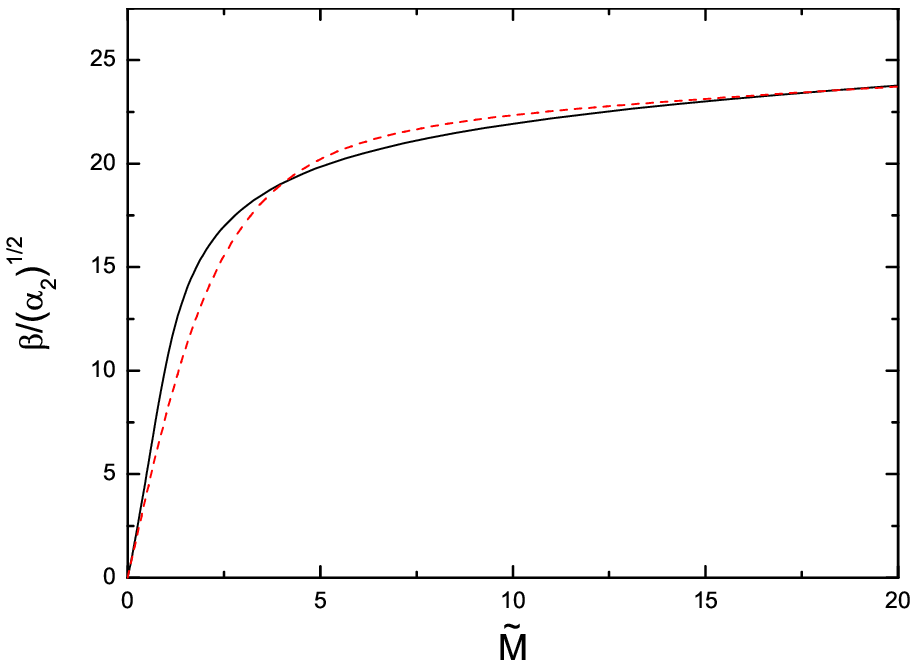}
\put(-112,-15){(c)}
\includegraphics[width=8cm]{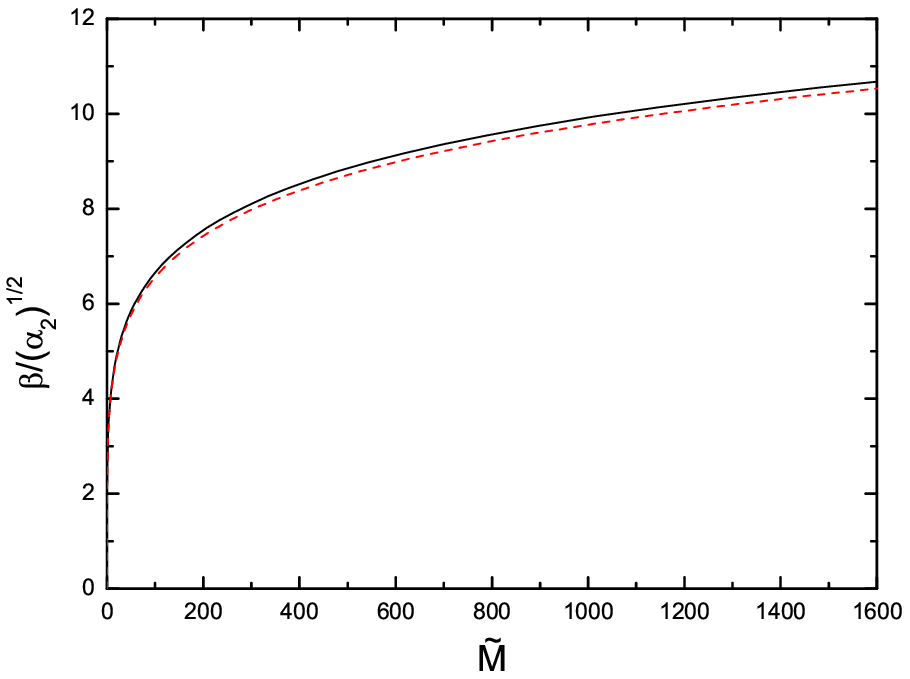}
\put(-110,-15){(d)}
\caption{The mass-temperature diagram in (a) four dimensions,
(b) five dimensions, (c) six dimensions, and  (d) ten dimensions,
where $\b \equiv 1/T$.
The dashed lines are the non-dilatonic case for comparison.}
\label{temperature}
\end{figure}

We show the $\tilde M$-$\beta$ relations in Fig.~\ref{temperature}, where
$\beta=1/T$ is the inverse temperature. In $D=4$, we can see that the GB term
has the tendency to raise the temperature compared to the non-dilatonic solution
(Schwarzschild black hole).
This behavior comes from the contribution of $\tm'_H$ of Eq.~(\ref{temp}).
$\tm'_H$ has the minimum value $\tm'_H=-\frac12$ by Eqs. (\ref{phip4}) and (\ref{mp4})
at the singular point. Since the temperature takes nonzero finite value for
all the mass range, we expect that the black hole will not stop emitting
radiation and continue evaporating until the solution reaches the minimum
mass solution denoted by the point C in Fig.~\ref{d4}(e).

In the five-dimensional case, the non-dilatonic black holes have interesting
thermodynamic property. The temperature increases as the mass of the black
hole becomes small for large black holes, which means that the heat capacity
is negative. Below the mass $\tilde{M}=2.976072$, however,
the temperature decreases as the mass becomes small. The sign of the heat capacity
changes at this mass. This behavior is qualitatively same as
the Reissner-Nordstr\"om black hole solution and can be classified to
the second order phase transition. As the black hole becomes small through
Hawking radiation, the temperature becomes extremely law, and the solution
cannot reach the singularity with zero horizon radius.
This is favorable feature from the point of view of cosmic censorship
hypothesis while more detailed analysis should be necessary for the definite answer.

By adding the dilaton field to the system, we find that the
thermodynamic properties change drastically. The heat capacity is negative
in all the mass range, and the temperature blows up at the singular solution.
This is due to the nontrivial coupling between the dilaton field and the GB
term and the resultant divergence of the dilaton field at the horizon.

For $D\geq 6$, the behavior of the temperature is qualitatively the same as that
in the non-dilatonic case. The dilaton field has tendency to lower the temperature
for the large black hole, while it raises the temperature for small black hole.
In six dimensions and higher ($D\geq 6$), our analysis shows that the temperature
diverges for the zero mass ``solution" and the black hole continues evaporating.

In GR, the horizon radius of the black hole is related to entropy
by $S=\pi r_H^2$. Hence the dependence of entropy on the mass $M$ can be easily
estimated from the $M$-$r_H$ diagram. In our case with GB gravity, however,
entropy is not obtained by a quarter of the area of the event horizon.
Along the definition of entropy in Ref.~\citen{Wald}, which originates from
the Noether charge associated with the diffeomorphism invariance of
the system, we obtain
\bea 
S=-2\pi \int_\Sigma \frac{\pa {\cal L}}{\pa R_{\mu\nu\rho\sigma}}
\e_{\mu\nu} \e_{\rho\sigma},
\ena 
where $\Sigma$ is the event horizon $(D-2)$-surface, ${\cal L}$ is the Lagrangian
density, $\e_{\mu\nu}$ denotes the volume element binormal to $\Sigma$.
This entropy has desirable properties such that it obeys the first law of black
hole thermodynamics and that it is expected to obey even the second law~\cite{Jacobson}.
For our present model, this gives
\bea 
S = \frac{A_H}{4}\left[1+2(D-2)_3 \frac{\a_2 e^{-\c\phi_H}}{r_H^2} \right],
\ena 
where $A_H==\frac{2 \pi^{(D-1)/2}}{\Gamma((D-1)/2)} r_H^{D-2}$ is
the area of the event horizon.

\begin{figure}
\includegraphics[width=8cm]{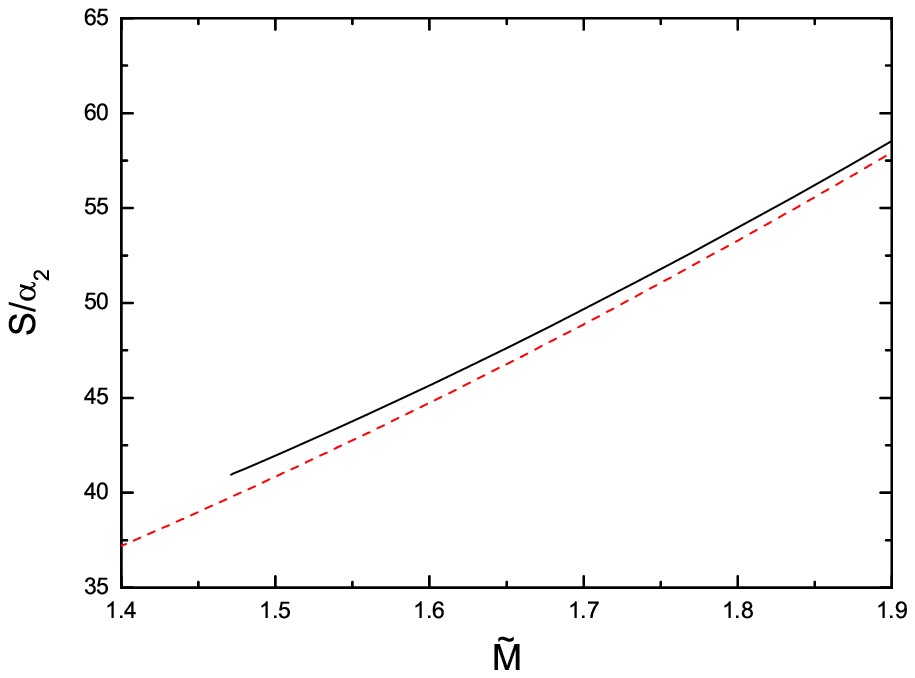}
\put(-112,-15){(a)}
\includegraphics[width=8cm]{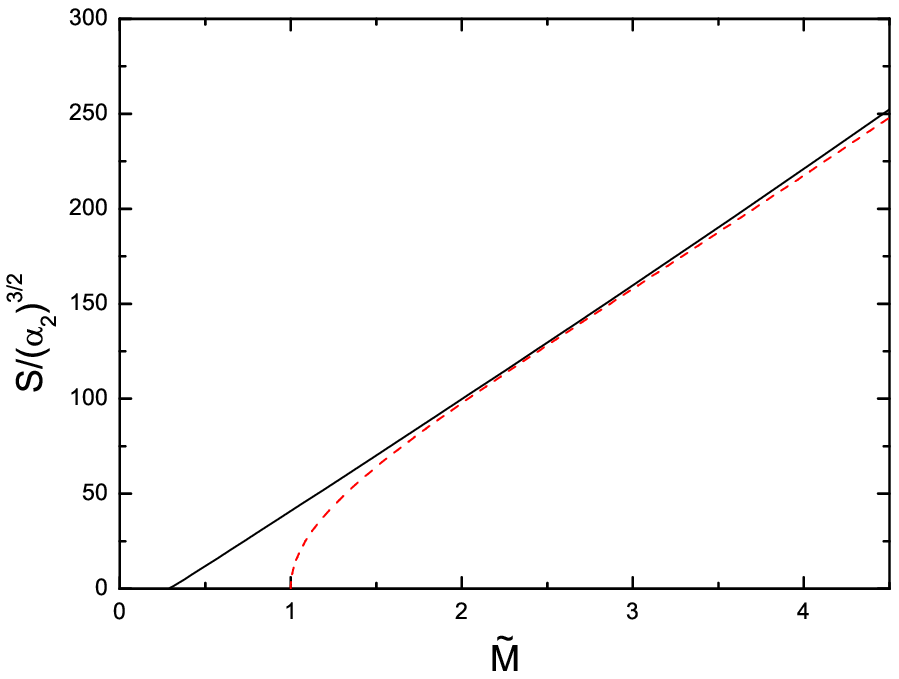}
\put(-110,-15){(b)}
\vspace*{1mm}\\
\includegraphics[width=8cm]{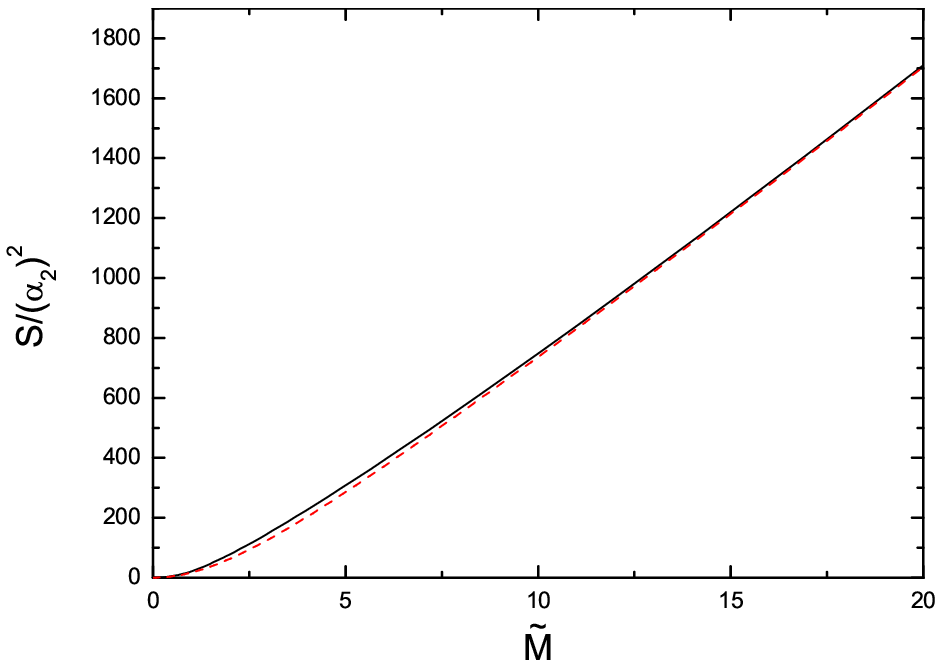}
\put(-112,-15){(c)}
\includegraphics[width=8cm]{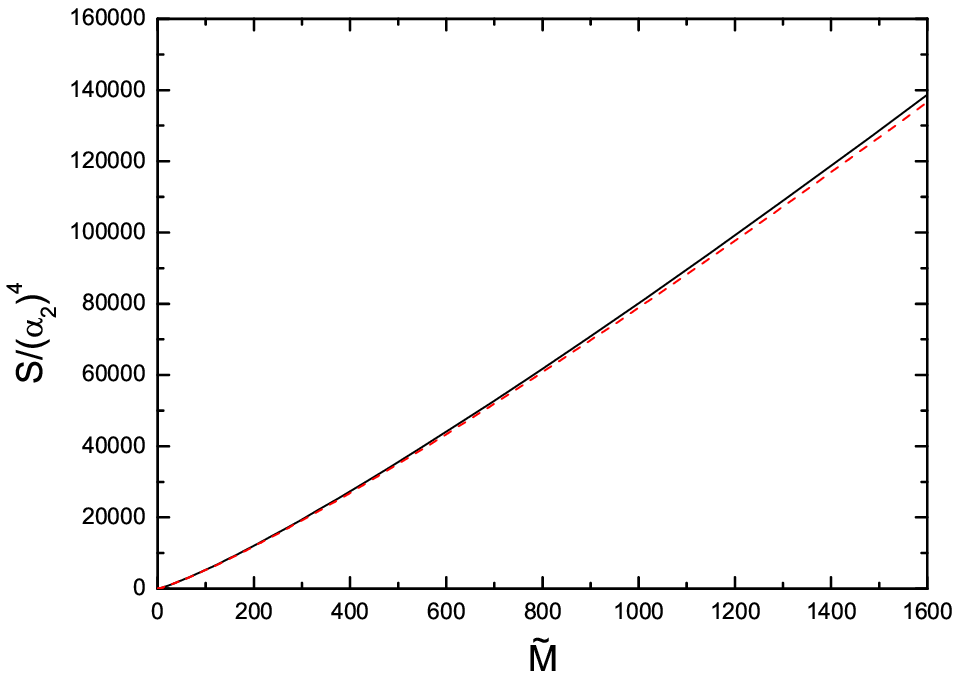}
\put(-112,-15){(d)}
\caption{The mass-entropy diagram in (a) four dimensions,
(b) five dimensions, (c) six dimensions, and  (d) ten dimensions.
The dashed lines are the non-dilatonic case for comparison.
}
\label{entropy}
\end{figure}

Fig.~\ref{entropy} shows the $M$-$S$ plots of our solutions. Although there
is no qualitative difference between the dilatonic and the non-dilatonic
cases except that the solution disappears at the nonzero finite mass for
$D=4$ (the dilatonic solution) and $D=5$,
it should be noted that entropy of the dilatonic black hole is always
larger than that of the non-dilatonic black hole with the same mass.
We have now no physical interpretation for this
behavior both in the classical and quantum levels since they are different systems.

\section{Conclusions and discussions}

We have studied black hole solutions in the dilatonic Einstein-Gauss-Bonnet theory
in $D$ dimensional spacetime. We assumed that spacetime is static and
spherically symmetric, and asymptotically flat. Our analysis is the direct
extension of the four-dimensional dilatonic system to higher dimensional
spacetime and of the non-dilatonic black holes to the dilatonic ones
in higher-dimensional solutions. Our black holes show remarkable properties,
which depend strongly on dimensions. We focused on the $D=4, 5, 6$ and $10$
cases in the numerical analysis, and compared the solutions with the known
non-dilatonic solution. We also studied their thermodynamic properties.

Firstly, in four dimensions, we made a consistency check on our solutions
with those previously investigated. Spacetime around the event horizon has
regions where the effective energy density becomes negative.
There is the minimum mass of black hole below which no regular solution exist.
There is also a solution which has minimum horizon-radius.

In five dimensions, we find that the effects of the GB term is negligible
for the large black holes ($r_H\gg \ell_s$), and the solution can be well
approximated by Tangherlini solution with the dilaton field that decays with power.
In contrast, for the small black holes ($r_H\leq \ell_s$), spacetime
is divided into the GR region and the GB region with a sharp transition.
In the GB region the dilaton field behaves logarithmically and the effective
energy density becomes negative. The regular black hole solutions exist for
all horizon radius. In the zero horizon-radius limit the solution becomes singular.
These properties are same as those of the non-dilatonic solutions while the mass
in this limit is different due to diverging energy density at the center.

In higher dimensions ($D\geq 6$), for the spacetime of the small black holes
($r_H \ll \ell_s$), the string effect extends to just around their event
horizons which are much smaller than the string scale. This is a remarkable
property since one naturally expects that the string effect extends to
$\ell_s$ in any situation. This is, however, not the case.
The regular solution exists for any horizon-radius. In the zero horizon-radius
limit, the mass of the solution approaches zero which is different from
the lower dimensional cases.

In the non-dilatonic case, the thermodynamic properties are similar for
all dimensions except for the five dimensions. In five dimensions,
the specific heat is negative for large black holes but it becomes positive
for small mass solutions through the second order phase transition.
For the dilatonic black holes, all of them have negative
specific heat without phase transition and have nonzero temperature. This suggests
that the black holes continue evaporating to the singular (in $D=4$ and $5$)
and zero mass solutions (in $D\geq 6$). Entropies of our solutions are larger
than that of the non-dilatonic solutions.

There are some related issues to our solutions.
All the solutions that we have obtained in this paper correspond to
the GR branch in the non-dilatonic solutions. We have found that there is no black hole
solution in the non-GR branch for $\alpha_2 > 0$.
Since our numerical analysis was limited to outer spacetime of
the event horizon, the global structures of the solutions such as
the existence of the inner horizon and (central or branch) singularity
have not been clarified. This may be done by integrating field equations
inward numerically.

The ambiguity of the frames is also important. In this paper, we have studied
in the Einstein frame. There is, however, a possibility that the properties
of solutions changes drastically by transforming to the string frame.
In particular, the conformal transformation may become singular.

For the choice $\c=\sqrt{2/(D-2)}$,
\footnote{
The following discussions go through for other choice of $\c$.}
the system in the string frame
\bea 
S=\frac{1}{2\kappa_D^2}\int d^Dx \sqrt{-\hat{g}}\; e^{-2\phi}
\left[\hat{R} + 4 (\partial_\mu \phi)^2
 + \a_2 \hat{R}^2_{\rm GB} \right],
\ena 
is obtained by the conformal transformation~\cite{BGO}
\bea
\hat{g}_{\mu\nu}=\Omega^2 g_{\mu\nu};~~~~
\Omega = e^{-\frac{\gamma}{2}\phi}.
\ena 
The metric in the string frame becomes
\bea 
\d\hat{s}^2=-Be^{-2\delta-\gamma\phi}dt^2
+B^{-1}\biggl(1-\frac{\gamma}{2}r\phi'\biggr)^{-2}d\hat{r}^2
+\hat{r}^2h_{ij}dx^jdx^i,
\ena 
where the prime denotes the $r$-derivative in the Einstein frame and
\bea
\hat{r}=re^{-\frac{\gamma}{2}\phi}.
\ena
The difference between the frames appears for the small black holes since
the dilaton field behaves nontrivially.
In four dimensions, both $\phi_H$ and $\phi'_H$ are finite for
all solutions (although $\phi''$ diverges for the solution at the singular point).
This means that there is no qualitative difference in both frames.
In $D\geq 6$, both the $\phi_H$ and $\phi'_H$ vanish in the zero horizon-radius
limit. Hence the spacetime structures in both frames is exactly the same.
In five dimensions, $\phi_H\sim -2 \log \tr_H /\gamma$ and $\phi'_H\sim 1/\tr_H$
in zero horizon-radius limit.
Then, the prefactor of $\hat{g}_{rr}$, i.e., $(1-\gamma r\phi'/2)^{-2}$ takes
nonzero finite value, and the metric does not give throat like structure but
ordinary black hole spacetime unlike the extreme solution
in the Einstein-Maxwell-dilaton system~\cite{GHS}.

The black hole solutions can be applied to brane world cosmology.
By adopting our solutions to the bulk solution, we may obtain the dilatonic
braneworld with the GB term. This is the dilatonic extension of the vacuum
analysis\cite{MT}. Since the dilaton field shows nontrivial behavior near
the event horizon, the early evolution of cosmology would be affected strongly.

The stability of our solutions is another important subject to study~\cite{TYM,torii,MS}.
It would be also interesting to extend our analysis to dilatonic black holes
(large and small) with charges.\cite{CCMOP}

We hope to report results on these issues as well as topological black holes elsewhere.

\section*{Acknowledgements}
The work of Z.K.G. and N.O. was supported in part by the Grant-in-Aid for
Scientific Research Fund of the JSPS Nos. 20540283 and 06042.
The work of N.O. was also supported by the Japan-U.K. Research Cooperative Program.

\end{document}